\documentclass[12pt]{article}
\pdfoutput=1        
    
\usepackage{hyperref}      

\usepackage{amsmath}       
\usepackage{amssymb}  
\usepackage{amsthm} 

\usepackage{graphicx} 
\usepackage{url} 
\usepackage{enumerate} 
\usepackage{color}
\usepackage{ulem}
\usepackage{float}
\usepackage{wrapfig} 

\allowdisplaybreaks[1]
\def\eq#1{(\ref{#1})}
\def\s[#1\s]{\begin{align}\begin{split}#1\end{split}\end{align}}
\def\[#1\]{\begin{align}#1\end{align}}

\def\tphi{\tilde \phi}
\def\bphi{\bar \phi}
\def\tphi{\tilde \phi}
\def\teta{\tilde \eta}
\def\pp{{\perp\hspace{-.1cm}\perp}}
\def\blam{\bar \lambda}

\begin{document}

\begin{titlepage} 

\title{
\hfill\parbox{4cm}{ \normalsize YITP-25-48}\\   
\vspace{1cm} 
Renormalization group-like flows\\
in randomly connected tensor networks 
}

\author{Naoki Sasakura\footnote{sasakura@yukawa.kyoto-u.ac.jp}
\\
{\small{\it Yukawa Institute for Theoretical Physics, Kyoto University, }}\\
{\small {\it and } } \\
{\small{\it CGPQI, Yukawa Institute for Theoretical Physics, Kyoto University,}} \\
{\small{\it Kitashirakawa, Sakyo-ku, Kyoto 606-8502, Japan}}
}
 
\date{\today} 
 
%\date{April 21, 2023}

\maketitle 

\begin{abstract}
Randomly connected tensor networks (RCTN) are the dynamical systems defined by summing over all the possible networks of tensors.
Because of the absence of fixed lattice structure, RCTN is not expected to have renormalization procedures.
In this paper, however, we consider RCTN with a real tensor, 
and it is proven that a Hamiltonian vector flow of a tensor model in the canonical formalism with a positive cosmological
constant has the properties which a renormalization group (RG) flow of RCTN would have: 
The flow has fixed points on phase transition surfaces; 
every flow line is asymptotically terminated by fixed points at both ends, where 
an upstream fixed point has higher criticality than a downstream one;
the flow goes along phase transition surfaces;
there exists a function which monotonically decreases along the flow, analogously to the $a$- and $c$-functions of RG. 
A complete classification of fixed points is given. Although there are no cyclic flows in the strict sense, 
these exist, if infinitesimal jumps are allowed near fixed points.
\end{abstract}
\end{titlepage}  
 
\section{Introduction}
\label{sec:introduction}
While general relativity and quantum mechanics are the two established foundations of modern physics, they suffer from the difficulty
of unification, namely, of constructing quantum gravity. 
This problem is not only for theoretical consistency, but is also of observational interest, since quantum gravity is needed to study some basic 
astrophysical questions, such as fates of black holes, birth of universe/spacetime, and so on.  
An attractive direction toward quantum gravity is to abandon the continuous spacetime notion
as a foundation, and rather consider that continuous spacetimes emerge in macroscopic scales from microscopic discrete structures.
This idea would be in accord with various spacetime uncertainties proposed by qualitative estimates of 
quantum gravitational effects \cite{Hossenfelder:2012jw,Sasakura:1999xp}. 
In fact there are various approaches to quantum gravity with different microscopic discrete structures, 
such as Regge calculus \cite{Regge:1961px}, loop quantum gravity \cite{Rovelli:2004tv}, 
causal sets \cite{Surya:2019ndm}, dynamical triangulations \cite{Loll:2019rdj}, 
matrix model \cite{Eynard:2016yaa}, tensor model \cite{Ambjorn:1990ge,Sasakura:1990fs,Godfrey:1990dt,Gurau:2009tw}, 
quantum graphity \cite{Konopka:2006hu}, and so on. 
To the author's knowledge, there are so far no truly successful theories with emergent continuous spacetimes. 

In this paper we specifically consider a system with tensor networks. 
The tensor network method is the new developing technique which can be applied
to solving various quantum many body systems \cite{Orus:2013kga}.
An interesting fact is that there are tensor renormalization procedures \cite{Levin:2006jai,Xie:2012mjn}, and, 
by repeatedly applying them, network structures can be made unlimitedly smaller.
Therefore in tensor networks the emergence of continuum spacetimes is realized by the presence of tensor renormalization procedures.

The tensor networks considered above assume certain macroscopic arrangements of networks to 
approximate some continuous spacetimes in discrete manners.  
From the view point of quantum gravity, however, such macroscopic arrangements
are not appropriate, since this assumes certain pregeometric structures. Instead a more natural formulation for quantum gravity is to 
allow all the possible networks of tensors. More exactly, we are interested in systems in which all the possible 
networks of tensors are summed up. We call it randomly connected tensor networks (RCTN). 
RCTN can describe various physical systems on random networks, which have extensively been studied in the literature 
\cite{Bachas:1994qn,dembo,dembo+,Dorogovtsev:2002ix,Dorogovtsev:2008zz,johnston,leone,Sasakura:2014zwa,Sasakura:2014yoa}.

An interesting possibility of connection between RCTN and quantum gravity was found in \cite{Narain:2014cya}. 
Here the model of quantum gravity is what we call the canonical tensor model (CTM) \cite{Sasakura:2011sq,Sasakura:2012fb}.  
It is a tensor model in the canonical formalism and mimics the structure of the Arnowitt-Deser-Misner formalism of general relativity
\cite{Arnowitt:1962hi,DeWitt:1967yk,Hojman:1976vp}.  
Generally, in the canonnical formalism of quantum gravity, one needs to obtain a wave function of the ``universe" by solving 
the Wheeler-DeWitt equation \cite{DeWitt:1967yk}. 
In the case of CTM the solution has the expression of summing over all the ways of connections of tensors, 
when it is formally expanded in perturbations of tensors.

The above finding motivated a full 
understanding of the connection between RCTN and CTM. When the tensors in RCTN are real, RCTN can be regarded as 
classical statistical systems. In \cite{Sasakura:2014zwa,Sasakura:2015xxa}, the Hamiltonian vector flows 
using the Hamiltonian of CTM were interpreted as renormalization group (RG) flows in RCTN. However, the success was partial.
For instance, while there are some correlations between the flows and the phase structures,
some fixed points are not correlated with phase transitions. This is different from what we see in statistical systems. 

In this paper we establish the correct connection between RCTN and CTM: the flow defined by CTM has the properties 
an RG flow would have in RCTN. 
The essential difference from the previous studies is that we take into account the positive cosmological constant term 
of the Hamiltonian of CTM to define the flow by a Hamiltonian vector flow. A partial list of the successes is given below:
\begin{itemize}
\item
The flow has fixed points on phase transition surfaces.
\item
Every flow line is asymptotically terminated by fixed points at both ends, where
an upstream fixed point has higher criticality than a downstream 
one\footnote{More exactly, an upstream fixed point has larger 
$N_+$ than a lower one has, where $N_+$ will be defined in Section~\ref{sec:decompP}.}.
\item
The flow goes along phase transition surfaces.
\item
There exists a function which monotonically decreases along the flow, analogously to the $a$- and $c$-functions of 
RG \cite{Zamolodchikov:1986gt,Cardy:1988cwa} . 
\item
A complete classification of fixed points is given.
\end{itemize}

This paper is organized as follows. In Section~\ref{sec:defRCTN} we define RCTN which we study. We only consider the case that 
the tensor is a symmetric real tensor with three indices. In Section~\ref{sec:RCTN} we define the thermodynamic limit of RCTN, where
the size of networks is taken infinitely large. It is shown that the free energy can be computed by solving an equation for the minimum. 
In Section~\ref{sec:phaseRCTN} we discuss the critical points. In Section~\ref{sec:decompP} we study a decomposition of the tensor
by which the flow is represented in a convenient manner. 
In Section~\ref{sec:floweq} we define the flow equation of RCTN by a Hamiltonian vector flow using the Hamiltonian of CTM. 
In Section~\ref{sec:firstorder} we prove that the flow is along the first-order phase transition surfaces. 
In Section~\ref{sec:flow} we study the flow equation in terms of the decomposition of the tensor. We introduce a 
label which classifies the fixed points. 
In Section~\ref{sec:criticalexp} we study the flow near fixed points, and obtain the dimensions of the relevant, irrelevant, and 
marginal directions. The critical exponent is computed and agrees with what is expected from the mean field analysis.
In Section~\ref{sec:analog} we define an RG-function which monotonically decreases under the flow. 
This is an analogue of the $a$- and $c$- functions of quantum field theories. 
In Section~\ref{sec:deg} we discuss an ambiguity of the label on the first-order phase transition surfaces. 
This ambiguity allows the presence of cyclic flows\footnote{The possibility of cyclic RG flows was pointed out in \cite{Curtright:2011qg}.}, 
if infinitesimal jumps of the tensor are allowed near the fixed points. 
In Section~\ref{sec:identity} we derive the flow equation by taking the thermodynamic limit of 
an identity of the system. 
In Section~\ref{sec:examples} we give simple examples of our results.  
The last section is devoted to summary and future prospects.

\section{Randomly connected tensor network}
\label{sec:defRCTN}
In this section we define randomly connected tensor networks (RCTN).  
The tensor we consider is an order-three real symmetric tensor of dimension $N$: 
$P_{abc}\in \mathbb{R},\ P_{abc}=P_{bca}=P_{bac}\ (a,b,c=1,2,\ldots,N)$. 
The partition function of RCTN is defined by \cite{Sasakura:2014zwa,Sasakura:2014yoa,Sasakura:2015xxa}
\[
Z_n(P)=\frac{1}{n!}\int_{\mathbb{R}^N}
\frac{d^N\tilde\phi}{(2 \pi)^\frac{N}{2}} \left( \frac{1}{6} P\tilde\phi^3 \right)^n e^{-\frac{1}{2} \tilde\phi^2},
\label{eq:ZofRCTN}
\]
where we have used the following shorthand notations,
\[
\begin{split}
P\tilde \phi^3&\equiv P_{abc}\tilde \phi_a \tilde\phi_b \tilde\phi_c, \\
\tilde \phi^2&\equiv \tilde\phi_a \tilde\phi_a,
\end{split}
\]
and the integration is over the whole $N$-dimensional real space. Here $n$ is taken even, 
because \eq{eq:ZofRCTN} identically vanishes for odd $n$. 
We employ the convention that repeated indices are summed over,
unless otherwise stated. 

The system \eq{eq:ZofRCTN} has an orthogonal group symmetry:
\[
\tphi_a'=M_{ab} \tphi_{b},\ P'_{abc}=M_{ad}M_{be}M_{cf} P_{def}.
\label{eq:ontrans}
\]
where $M_{ab}$ belongs to the fundamental representation of the real orthogonal group $O(N,\mathbb{R})$.

\begin{figure}
\begin{center}
\includegraphics[width=3cm]{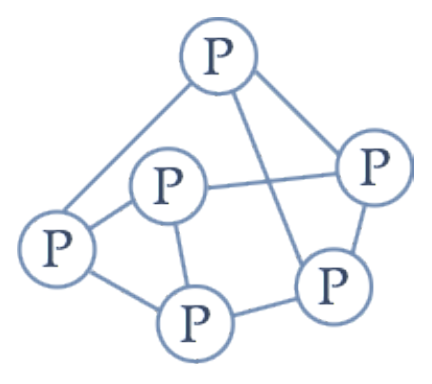}
\caption{An example of a network $g$ for $n=6$.
\label{fig:networkP}}
\end{center} 
\end{figure}

The Gaussian integration \eq{eq:ZofRCTN} has a diagrammatic expression \cite{Peskin:1995ev},
\[
Z_n(P)=\sum_{g\in {\cal G}_n} \frac{1}{S_g}  \overbrace{P_{...}P_{...}\cdots P_{...}}^{n},
\label{eq:Znetwork}
\]
where the summation is over all the possible networks of $n$ trivalent vertices ${\cal G}_n$.
In each network $g$ the tensor $P$ is assigned to vertices, and edges represent index contractions (See Fig.~\ref{fig:networkP}). 
$1/S_g$ denotes the symmetry factor of $g$ \cite{Peskin:1995ev}. 

Due to the expression \eq{eq:Znetwork}, we call the system randomly connected tensor networks. 
With choices of $P$, the system can describe various statistical systems on random networks
\cite{Bachas:1994qn,dembo,dembo+,Dorogovtsev:2002ix,Dorogovtsev:2008zz,johnston,leone,Sasakura:2014zwa,Sasakura:2014yoa}.

\section{Thermodynamic limit of RCTN}
\label{sec:RCTN}
The thermodynamic limit of RCTN is given by the infinite size limit of networks, namely $n\rightarrow \infty$.
In fact, the free energy of the system \eq{eq:ZofRCTN} can exactly be computed in the thermodynamic limit 
in terms of a mean field.

To see this, let us first perform a rescaling of the field, $\tilde\phi=\sqrt{2n}\phi$. We obtain
\[
\begin{split}
 Z_n(P)=C_{N,n} \int_{P\phi^3>0} d^N \phi \, e^{-n(\phi^2-\ln (P\phi^3))},
\end{split}
\label{eq:rescaleZ}
\]
where
\[
C_{N,n}= \frac{2 (2n)^\frac{N+3n}{2}}{6^n(2\pi)^\frac{N}{2} n! }.
\]
Here, for later convenience, we have restricted the integration region of $\phi$ to be
\[
P\phi^3 >0,
\label{eq:positive}
\]
and have multiplied a factor of 2, because the $P\phi^3<0$ region has the same contribution as the $P\phi^3 >0$ region for even 
$n$\footnote{The contribution from $P\phi^3=0$ can be ignored, since the integrand vanishes.}.

From the expression \eq{eq:rescaleZ}, one can see that, in the thermodynamic limit $n\rightarrow\infty$,
the steepest descent method\footnote{In this real valued case, the method is also called the Laplace method.}
can be applied, and the partition function is determined by the absolute minimum of 
\[
f(P,\phi)=\phi^2- \ln ( P\phi^3 )
\label{eq:defofpphi}
\]
with respect to $\phi$. Note that the absolute minimum with real $\phi$ always exists, because $f(P,\phi)$ is positive infinity at the boundaries, $P\phi^3=0$ and $\phi=\infty$.  
More explicitly,  the free energy per vertex in the thermodynamic limit is given by
\[
f(P)=-\lim_{n\rightarrow \infty} \frac{1}{n} \ln\left( \frac{Z_n(P)}{C_{N,n}}\right)=
\min_\phi f(P,\phi) = f(P,\bphi),
\label{eq:freeenergy}
\]
where $\bar \phi$ is the location of the absolute minimum (or one of the absolute minima), 
and the numerical factor $C_{N,n}$ has been removed from the definition of the free energy, 
since it is $P$-independent. In this paper we often call $\bar \phi$ the ground state. 

The minimum location $\bar \phi$ is one of the solutions to the stationary condition,
\[
\left.\frac{\partial f(P,\phi)}{\partial \phi_a} \right|_{\phi=\bar \phi}
=2\bar \phi_a-\frac{3 (P\bar \phi^2)_a}{P\bar \phi^3}  =0,
\label{eq:stationary}
\]  
where a shorthand notation,
\[
(P\phi^2)_a \equiv P_{abc} \phi_b \phi_c,
\]
is used. 
Note that the two-fold degeneracy $\bar\phi \leftrightarrow -\bphi$ of the solution to the stationary condition \eq{eq:stationary} is removed by 
the condition \eq{eq:positive}, namely, we are taking the solution satisfying
\[
P\bar \phi^3>0.
\label{eq:convphi}
\]
In later computations, we often use \eq{eq:stationary} in the form,
\[
(P\bar \phi^2)_a=\frac{2 (P\bar \phi^3)}{3}\bar \phi_a,
\label{eq:simple}
\]
to simplify expressions. As will be explained below, \eq{eq:simple} has the form of the eigenvalue/vector equation of a tensor. 
By multiplying $\bar \phi$ to \eq{eq:stationary}, we obtain
\[
\bar \phi^2=\frac{3}{2}.
\label{eq:phi2}
\]
Therefore, as will be derived below, the free energy \eq{eq:freeenergy} can alternatively be expressed as
\[
f(P)=\frac{3}{2}-\log ( P\bphi^3)=\frac{3}{2} -\log\left( \max_{\genfrac{}{}{0pt}{}{\phi\in \mathbb{R}^N}{|\phi|^2=3/2}} P \phi^3 \right),
\label{eq:newfree}
\]
where the norm is defined by $|\phi|=\sqrt{\phi_a \phi_a}$.

A comment is in order. The minimum solution $\bphi$ may have degeneracy for certain $P$. Even for such $P$, 
the expression \eq{eq:freeenergy} and therefore \eq{eq:newfree} are correct by freely taking any one of the solutions as $\bar \phi$.
This is because, as long as $N$ is finite, the effect of such degeneracy to the partition function \eq{eq:rescaleZ} 
is at most in a finite power of $n$, and can be ignored in \eq{eq:freeenergy}, because $\lim_{n\rightarrow \infty}\log n /n\rightarrow 0$.

RCTN has intimate connections to the tensor eigenvalue/vector problem \cite{Qi, lim, cart, qibook},
and the $p$-spin spherical model for spin glasses \cite{pspin,pedestrians}. A Z-eigenpair 
$(z, w)$ of a symmetric real order-three dimension-$N$ tensor $P$ is defined by a solution to 
\[
P_{abc} w_b w_c=z\, w_a ,\ |w|=1, z \in \mathbb{R},\ w \in \mathbb{R}^N,
\label{eq:eigen}
\]
where $z$ and $w$ are a real eigenvalue and a real eigenvector of a tensor $P$, respectively.
In fact the largest eigenvalue $z_\text{max}$ is related to the injective norm $|\cdot |_\text{inj}$ 
of a tensor $P$:
\[
|P|_\text{inj}\equiv
\max_{|w|=1} P w^3=z_\text{max}.
\label{eq:lammax}
\]
This can be proven by the method of Lagrange multiplier; considering the stationary condition for $\frac{1}{3} Pw^3-\frac{1}{2}z (w^2-1)$ 
by introducing a Lagrange multiplier $z$ for the constraint $|w|=1$, and seeing that $P w^3=z$ for the
solution. The last equation in \eq{eq:newfree} can be derived in a similar way for the normalization $|\phi|^2=3/2$. 
In addition, \eq{eq:lammax} also implies that $z_\text{max}$ and the corresponding eigenvector $w_\text{max}$
respectively give the ground state energy and the ground state for the $p$-spin spherical model, which has the Hamiltonian,
\[
H=-Pw^3, \ |w|=1,\ w\in \mathbb{R}^N.
\]

It has been proven that the tensor eigenproblem is NP-hard \cite{nphard}. Because of this, explicitly obtaining $\bphi$ is 
generally hard for large $N$.

\section{Critical points of RCTN}
\label{sec:phaseRCTN}
Critical points of RCTN are the branching loci of the minimum location $\bphi$, 
which is a solution to the stationary equation \eq{eq:stationary}. 
Therefore critical points are the locations where the Hessian matrix,
\[
K_{ab}=\left. \frac{1}{2} \frac{\partial^2 f(P,\phi)}{\partial \phi_a \partial \phi_b} \right |_{\phi=\bphi},
\label{eq:defofK}
\]
contains zero eigenvalues, while the other eigenvalues must be positive for the stability of the minimum.
By taking the derivatives and using \eq{eq:simple}, we obtain
\[
\begin{split}
K_{ab}&=\delta_{ab} -\frac{3 (P\bar \phi)_{ab}}{P\bar \phi^3}+\frac{9}{2} \frac{(P\bar \phi^2)_a 
(P\bar \phi^2)_b}{(P\bar \phi^3)^2}  \\
&=\delta_{ab} +2 \bar \phi_a \bar \phi_b -3 R_{ab}, 
\end{split}
\label{eq:Mab}
\]
where we have introduced 
\[
R_{ab}\equiv \frac{(P\bar\phi)_{ab}}{P\bar\phi^3}
\label{eq:defofR}
\]
with a shorthand notation,
\[
(P\phi)_{ab}\equiv P_{abc}\phi_c.
\]

From \eq{eq:simple} and \eq{eq:phi2}, one can find that 
$\bar \phi$ is an eigenvector of $K$ and $R$ with constant eigenvalues:
\s[
R\bar \phi&=\frac{2}{3} \bar\phi, \\
K \bar \phi&=2 \bar \phi,
\label{eq:Rphi}
\s]
where $(R\bphi)_a=R_{ab}\bphi_b$, and similarly for $K\bphi$.
Since $K,R$ are real symmetric matrices, the other eigenvectors are orthogonal to $\bar \phi$.
Let us decompose the index vector space $V$ into the parallel and the transverse 
subspaces against $\bphi$ as $V=V_\parallel \oplus V_\perp$, where $\oplus$ denotes the direct sum.
Then $R$ can be decomposed into those on each subspace,
\[
R=\frac{4}{9}  \bar \phi \otimes \bar\phi+R^\perp,
\label{eq:Rdecomp}
\]
where $\otimes$ denotes the tensor product $(\bphi\otimes \bphi)_{ab}= \bphi_a \bphi_b$, 
$R^\perp \in [V_\perp \otimes  V_\perp]$,
and the numerical factor of the first term is due to \eq{eq:phi2} and \eq{eq:Rphi}. 
Here we have introduced a notation: the square brackets $[\cdot]$ represent symmetrization with respect to the indices,
and $R^\perp \in [V_\perp \otimes  V_\perp]$ means that $R^\perp_{ab} v^1_a v^2_b \neq 0$ only if $v^1,v^2\in V_\perp$.
The same notation will also be used for tensors in later discussions.

Since
\[
K^\perp=I^\perp-3 R^\perp
\label{eq:kr}
\]
from \eq{eq:Mab}, where $I^\perp$ is the projection onto $V_\perp$, and
the eigenvalues of $K$ must be non-negative for the stability of the minimum location, 
all the eigenvalues $e^\perp_i\ (i=1,2,\ldots,N-1)$ of $R^\perp$ must satisfy
\[
e^\perp_i \leq \frac{1}{3}.
\label{eq:stability}
\] 
Note also that
\[
{}^\exists e^\perp_i=\frac{1}{3} \longleftrightarrow \bphi\hbox{ is a critical point},
\label{eq:criticaleg}
\] 
because $K$ then has the zero eigenvalue. 

\section{Decomposition of $P$}
\label{sec:decompP}
In this section we will discuss a decomposition of $P$ for further analysis. As proven in \ref{app:genP1}, $P$ has necessarily the form,
\[
P=\frac{8 (P\bphi^3)}{27}\bphi\otimes \bphi \otimes \bphi + 2 (P\bphi^3) \left[\bphi\otimes R^\perp\right]+P^\perp,
\label{eq:genP1}
\]
where the symmetrization of tensor $[\cdot]$, which was introduced below \eq{eq:Rdecomp} for matrices, is explicitly given by
\[
 \left[\bphi\otimes R^\perp\right]_{abc}= \frac{1}{3} \left( \bphi_a R^\perp_{bc}+\bphi_b R^\perp_{ca}
 +\bphi_c R^\perp_{ab}\right),
 \label{eq:symmetrization}
\]
and $P^\perp \in [V_\perp\otimes V_\perp\otimes V_\perp]$, namely, $P^\perp$ is a symmetric tensor and 
$P^\perp_{abc} v_a^1 v_b^2 v_c^3\neq 0$ only if $v^1,v^2,v^3 \in V_\perp$.

We will prove that the eigenvalues $e^\perp_i$ of $R^\perp$ have a lower bound in addition to the upper bound \eq{eq:stability}. 
Let us denote one of the eigenvalues 
of $R^\perp$ as $e$ and the corresponding eigenvector as $\eta_\perp \ (\eta_\perp\in V_\perp, \ |\eta_\perp|=1)$. 
Then consider a linear combination, 
$\tphi_\theta=\bphi \cos \theta +\eta_\perp |\bphi| \sin \theta$, where $|\tphi_\theta|^2=3/2$ because of \eq{eq:phi2}.
Then, from \eq{eq:genP1}, we obtain
\[
P\tphi_\theta^3=P\bphi^3 \left( 
\cos^3\theta +\frac{9e}{2}\cos \theta \sin^2 \theta +\frac{|\bphi|^3 \sin^3 \theta}{P\bphi^3} P^\perp \eta_\perp^3
\right).
\label{eq:pthetatilde}
\]
Since $P\phi^3\leq P\bphi^3$ for $^\forall \phi$ with $|\phi|^2=3/2$ from \eq{eq:newfree}, we in particular have 
$P\tphi_{\theta=2 \pi/3}^3+P\tphi_{\theta=4 \pi/3}^3\leq 2 P\bphi^3$. By explicitly computing this inequality by using \eq{eq:pthetatilde}, 
we obtain $e\geq-2/3$. Combining this with \eq{eq:stability}, we conclude
\[
-\frac{2}{3}\leq e^\perp_i \leq \frac{1}{3},
\label{eq:eineq}
\] 
for all the eigenvalues $e^\perp_i$ of $R^\perp$. 
The violation of the bound would contradict that $\bphi$ is the location of the absolute maximum of $P\phi^3$ (See \eq{eq:newfree}). 
The bound is in fact tight, as will be discussed in Section~\ref{sec:flow}. 

To further constrain the form of $P$ for later discussions, let us denote the degeneracies of the eigenvalues $1/3$ and $-2/3$ of $R^\perp$ as 
$N_+$ and $N_-$, respectively, and introduce $N_{\pp}=N-1-N_+-N_-$. 
From \eq{eq:eineq}, the other eigenvalues must satisfy $-2/3 < e^\pp_i<1/3\ (i=1,2,\cdots,N_\pp)$.
Now the transverse subspace $V_\perp$ is decomposed into a direct sum
$V_\perp=V_+\oplus V_-\oplus V_\pp$, where $V_+,V_-,V_\pp$ denote
the eigenvector subspaces corresponding the the eigenvalues, $1/3$, $-2/3$, and the others, respectively.  
Accordingly, $R^\perp$ can be decomposed into the form,
\[
R^\perp=\frac{1}{3} I^+ - \frac{2}{3} I^-+R^\pp,
\label{eq:rdecomp}
\]
where $I^+,I^-$ are the projections onto $V_+,V_-$, respectively, and $R^\pp \in [V_\pp\otimes V_\pp]$.
%Any of $I^+,I^-,R^\pp$ can be null, if any of $N_+,N_-,N_\pp$ vanishes, respectively. 
Putting \eq{eq:rdecomp} into \eq{eq:genP1}, we obtain
\[
P=\frac{8 (P\bphi^3)}{27}\bphi\otimes \bphi \otimes \bphi + \frac{2(P\bphi^3)}{3} \left[\bphi\otimes I^+\right]
- \frac{4(P\bphi^3)}{3} \left[\bphi\otimes I^-\right]+ 2(P\bphi^3) \left[\bphi\otimes R^\pp\right]+P^\perp.
\label{eq:pdecomp}
\]
 
As performed for $R^\perp$, $P^\perp$ can also be decomposed into the sum of $_5 C_2=10$ terms\footnote{A difference of 
$P^\perp$ from $R^\perp$ is that, while
$R^\perp$ is decomposed into a diagonal form in terms of the eigenvector subspaces of $R^\perp$, 
the decomposition of $P^\perp$ generally contains mixed ones like $P^{++-}$, and so on.} 
as $P^\perp=P^{+++}+P^{++-}+P^{++\pp}+P^{+--}+\cdots$, where 
$P^{ijk} \in [V_i\otimes V_j \otimes V_k]$\footnote{More explicitly, 
$P$ is a symmetric tensor, and $P^{ijk}_{abc} v_a^1 v_b^2 v_c^3\neq 0$, only if $v^1\in V_i, v^2\in V_j, v^3\in V_k$
or alternated cases.}.
Note that $P^{ijk}$ does not depend on the order of $i,j,k$ as notations.
As proven in \ref{app:pvanish} using similar discussions as above,
four of $P^{ijk}$ must vanish,
\[
P^{+++}=P^{--+}=P^{---}=P^{--\pp}=0.
\label{eq:pvanish}
\] 
Therefore, the general form of $P^\perp$ is given by
\[
P^\perp=P^{++-}+P^{++\pp}+P^{+-\pp}+P^{+\pp\pp}+P^{-\pp\pp}+P^{\pp\pp\pp}.
\label{eq:pperpgen}
\]

Let us summarize what we have obtained in this section. 
\begin{itemize}
\item
For general $P$, the original vector space $V$ of the index can be decomposed 
into $V=V_{\parallel}\oplus V_+\oplus V_- \oplus V_\pp$ according to the eigenvalue subspaces of $R$. 
Then $P$ has the decomposition \eq{eq:pdecomp}, where $I^+$ and $I^-$ 
are the projection to $V_+$ and $V_-$, respectively, and $R^\pp \in [V_\pp\otimes V_\pp]$. $R^\pp$ has eigenvalues in the 
range $-2/3 < e^\pp_i < 1/3$. $P^\perp$ is restricted to have the form \eq{eq:pperpgen}, where $P^{ijk} \in [V_i \otimes V_j\otimes V_k]$. 
\end{itemize}
 
\section{Flow equation}
\label{sec:floweq}
The flow equation we employ comes from a tensor model in the canonical formalism, which we call the canonical tensor model (CTM) \cite{Sasakura:2011sq,Sasakura:2012fb}. 
CTM has its motivation from quantum gravity. 
CTM classically has a similar structure as the Arnowitt–Deser–Misner (ADM) formalism
\cite{Arnowitt:1962hi,DeWitt:1967yk,Hojman:1976vp} of general relativity,
having the first-class constraints corresponding to the Hamiltonian and momentum constraints of ADM. 

The explicit form of the Hamiltonian constraint of the classical-mechanics model of CTM is given by \cite{Sasakura:2011sq,Sasakura:2012fb}
\[
{\cal H}_a=P_{abc}P_{bde}M_{cde}-\lambda M_{abb},
\label{eq:hamiltonian}
\]
where $M$ and $P$ are the real symmetric order-three tensors serving as the dynamical variables of CTM, and $\lambda$ is a real constant.
Here $M$ and $P$ are canonical conjugate to each other, satisfying the fundamental Poisson brackets, 
\s[
&\{ M_{abc}, P_{def} \} =\frac{1}{6}\sum_{\sigma} \delta_{a\,\sigma_d}\delta_{b\,\sigma_e}\delta_{c\,\sigma_f}, \\
&\{ M_{abc}, M_{def} \} =\{ P_{abc}, P_{def} \} =0,
\label{eq:fundpoi}
\s]
where the summation over $\sigma$ is over all the permutations of $d,e,f$, assuring the consistency of $M,P$ being symmetric tensors. 
We call $\lambda$ a cosmological constant, because the $N=1$ case of CTM 
agrees with the mini-superspace approximation of GR with a cosmological constant
proportional to $\lambda$ \cite{Sasakura:2014gia}.

To consider a Hamiltonian vector flow using \eq{eq:hamiltonian},
a flow direction $\varphi$ must be specified to construct a Hamiltonian by $\varphi_a {\cal H}_a$. 
In a former attempt \cite{Sasakura:2015xxa}  it has been argued that 
$\varphi=\bar \phi$ should be taken for the flow to be consistent with the phase structure. 
However, some of the fixed points of the flow were not correlated with the phase structure.
The difference of this paper from the former attempt is the inclusion of the cosmological constant term in \eq{eq:hamiltonian},
which was put zero in the former attempt.

The Hamiltonian \eq{eq:hamiltonian} contracted with the vector $\bar \phi$ generates a flow of $P$ given by
\[
\frac{d }{ds}P_{abc}=\{\bar \phi_d {\cal H}_d, P_{abc} \}=[\bar \phi P P]_{abc} -\lambda \, [\bar \phi \otimes I]_{abc},
\label{eq:Pflow}
\]
where $s$ parameterizes the trajectory of the flow, $I_{ab}=\delta_{ab}$, and 
\s[
[\bar \phi P P]_{abc}&=\frac{1}{3}\left( \bar \phi_d P_{dae}P_{ebc}+\bar \phi_d P_{dbe}P_{eca}+\bar \phi_d P_{dce}P_{eab}\right), \\
[\bar \phi \otimes I]_{abc}&= \frac{1}{3}\left( \bar \phi_a \delta_{bc}+\bar \phi_b \delta_{ca}+\bar \phi_c \delta_{ab} \right).
\s]
We take $\lambda$ to be a positive value, 
\[
\bar \lambda=\frac{8}{27}(P\bar\phi^3)^2.
\label{eq:lamval}
\]
Since this depends on $P$, it does not seem valid to regard it as a constant. 
However, we will prove that $\bar \lambda$ is in fact constant along the flow, justifying treating it as a constant in the discussions.

To prove the constancy, let us first discuss the flow of $\bphi$, which is determined by the flow of $P$. 
Since $\bar \phi$ is determined by the stationary condition \eq{eq:stationary},
the flow equation of $\bar\phi$ can be derived from the following consistency condition, 
\[
0=\frac{d}{ds} \frac{\partial f(P,\bar \phi)}{\partial \bar \phi_a}=
\frac{\partial^2 f(P,\bar \phi)}{\partial \bar \phi_a\partial \bar \phi_b} \frac{d \bar \phi_b}{ds}+
\frac{\partial^2 f(P,\bar \phi)}{\partial \bar \phi_a\partial P_{bcd}} \frac{d P_{bcd}}{ds}.
\label{eq:genflowphi}
\]
As proven in \ref{app:compsecond}, the second term on the right-hand side identically vanishes for the flow \eq{eq:Pflow}.   
Therefore, this uniquely determines 
\[
\frac{d \bar \phi}{ds}=0
\label{eq:dphids}
\]
almost everywhere except on the critical loci, where $K$ has zero eigenvalues
(See Section~\ref{sec:phaseRCTN}). 
Even on the critical loci, however, \eq{eq:dphids} is a consistent solution to \eq{eq:genflowphi}. 
Therefore we can consistently take \eq{eq:dphids} as the flow equation of $\bphi$ for the whole region. 

The above derivation of \eq{eq:dphids} based on \eq{eq:genflowphi} assumes that the same branch of the solution $\bphi$ 
to the stationary condition \eq{eq:stationary} continues to be taken under continuous change of $P$.  
However, this is not true on first-order phase transition surfaces, where there are transitions of branches.
In Section~\ref{sec:firstorder}, however, we will prove that the flows can never cross first-order phase transition surfaces,
validating the assumption.

A comment is in order. At certain $P$, $\bphi$ has degeneracy. In such a case, we may take one $\bphi$, and 
consider a flow line with $\bphi$ being kept constant, as required by \eq{eq:dphids}.  
In other words, the value of $\bphi$ to be taken on such $P$ generally depends on the flow line considered.
As far as it is kept constant along a flow line, the discussions in this paper are kept correct.

Along the flow, \eq{eq:Pflow} and \eq{eq:dphids}, we can prove the constancy of the positive cosmological constant $\bar \lambda$ 
in \eq{eq:lamval}:
\[
\begin{split}
\frac{d}{ds} (P\bar \phi^3)&=\frac{d P}{ds} \bar \phi^3 =(P\bar \phi^2)^2-\bar \lambda (\bar \phi^2)^2=0,
\end{split}
\label{eq:dpphi3ds}
\]
where we have used \eq{eq:simple} and \eq{eq:phi2}.  The constancy of \eq{eq:lamval} also leads to the constancy of the free energy 
along the flow, 
\[
\frac{d}{ds} f(P,\bar \phi)=0,
\label{eq:fconst}
\]
because of \eq{eq:newfree}.

\section{Flow on first-order phase transition surfaces}
\label{sec:firstorder}
In this section we will prove that the flow goes along the first-order phase transition surfaces.  This particularly implies 
that the flow cannot cross the first-order phase transition surfaces. 

First-order phase transition surfaces are the loci of $P$, where there are degeneracies of the ground state $\bphi$
of the free energy $f(P,\phi)$. They belong to the different branches of the solutions to 
the stationary condition \eq{eq:stationary}. 
Let us consider two of them, $\bphi_1$ and $\bphi_2$.
They satisfy
\[
\begin{split}
\bar \phi_1 & \neq \bar \phi_2, \\
f(P,\bar \phi_1)& =f(P,\bar \phi_2),  \\ 
\frac{\partial f(P, \bar \phi_1)}{\partial \bar \phi_{1a}} &=
\frac{\partial f(P, \bar \phi_2)}{\partial \bar \phi_{2a}}=0.
\end{split}
\label{eq:firstpoints}
\]
From \eq{eq:newfree} and \eq{eq:firstpoints}, we obtain
\[
P\bar\phi_1^3=P\bar\phi_2^3.
\label{eq:pphi312}
\]
 
Let us consider the flow \eq{eq:Pflow} with $\bar \phi=\bar \phi_1$ and the positive cosmological constant \eq{eq:lamval},
\[
\bar \lambda_1=\frac{8}{27}(P\bar\phi_1^3)^2.
\label{eq:lam1}
\]
As shown in \eq{eq:fconst}, we have
\[
\frac{d}{ds_1} f(P,\bar\phi_1)=0,
\label{eq:dfphi1zero}
\]
where we have used the notation $s_1$ for the parameter along the flow line
to stress that the flow direction is determined by $\bphi_1$, but 
not by $\bphi_2$.    On the other hand,
\[
\begin{split}
\frac{d}{ds_1} f(P,\bar \phi_2)
&=\frac{\partial  f(P,\bar \phi_2)}{\partial P_{abc}}\frac{d P_{abc}}{ds_1}+
\frac{\partial  f(P,\bar \phi_2)}{\partial \bar\phi_{2a}}\frac{d \bar\phi_{2a}}{ds_1} \\
&=-\frac{\bphi_{2a} \bphi_{2b}\bphi_{2c}}{P\bar\phi^3_2}
\left(
[\bar \phi_1 P P]_{abc} -\bar \lambda_1 [\bar \phi_1 \otimes I]_{abc}
\right) \\
&=-\frac{1}{P\bar\phi^3_2}
\left(
(P\bar\phi_1 \bar\phi_2)_a (P\bar\phi_2^2)_a - \bar \lambda_1 \bar \phi_2^2(\bar \phi_{1}\cdot \bar \phi_{2}) 
\right) \\
&=- \frac{1}{P\bar\phi^3_2}
\left(
\left(\frac{2 P\bar \phi_2^3}{3}\right)^2 (\bar \phi_1 \cdot \bar \phi_2)- \frac{3}{2}\bar \lambda_1 (\bar \phi_{1}\cdot \bar \phi_{2}) 
\right) \\
&=0,
\label{eq:dfphi2ds}
\end{split}
\] 
where we have used \eq{eq:simple} twice and \eq{eq:phi2} from the third to the fourth lines, 
and have finally put \eq{eq:pphi312} and \eq{eq:lam1}. 
Note that, from the first to the second lines, the value of $\frac{d \bar\phi_{2a}}{ds_1}$ does not matter 
because of the last equation of \eq{eq:firstpoints}.
Thus, combining with \eq{eq:dfphi1zero}, 
we have proven that $f(P,\bar \phi_1)=f(P,\bar\phi_2)$ is kept along the flow determined by $\bphi_1$. By exchanging
$\bphi_1$ and $\bphi_2$, the same holds for  the flow determined  by $\bphi_2$.
Thus, we have shown 
\begin{itemize}
\item
If a flow line contains a point on a first-order phase transition surface, the whole flow line is contained on the surface.  
In particular, the flow does not cross the first-order phase transition surfaces.
\end{itemize}

\section{Analysis of the flow and its fixed points}
\label{sec:flow}
In this section we will discuss more details of the flow equation \eq{eq:Pflow} by taking advantage of the decomposition
obtained in Section~\ref{sec:decompP}. By using \eq{eq:genP1}, \eq{eq:dphids} and \eq{eq:dpphi3ds}, 
the left-hand side of \eq{eq:Pflow} is given by
\[
\frac{d}{ds}P=2 (P\bphi^3)\, \left[ \bphi\otimes \frac{d R^\perp}{ds}\right]+\frac{d}{ds} P^\perp.
\label{eq:leftdp}
\]

As for the right-hand side of \eq{eq:Pflow}, by using \eq{eq:phi2}, \eq{eq:defofR}, \eq{eq:Rdecomp}, \eq{eq:genP1}, \eq{eq:lamval},
and $I=\frac{2}{3} \bphi \otimes \bphi+I^\perp$, we obtain 
\s[
[\bar \phi P P] -\blam \, [\bar \phi \otimes \hbox{I}]&=(P\bphi^3) [RP]-\blam [\bphi\otimes I] \\
&= \left( \frac{4 P\bphi^3}{9}\right)^2 \bphi\otimes \bphi\otimes \bphi +\frac{4}{9} (P\bphi^3)^2 [\bphi\otimes R^\perp]+
P\bphi^3 [R^\perp P]-\blam [\bphi\otimes I] \\
&= \frac{4 (P\bphi^3)^2}{3} \left[ \bphi \otimes 
\left(R^{\perp\,2} +\frac{1}{3} R^\perp -\frac{2}{9} I^\perp
\right)
\right] +P\bphi^3 [R^\perp P^\perp],
\s]
where the product of a matrix and a tensor is defined by $(RP)_{abc}\equiv R_{ad}P_{dbc}$.   
Note that the $\bphi\otimes \bphi \otimes \bphi$ term in the second line has been canceled with a part of the last term.
Comparing this with \eq{eq:leftdp}, we obtain
\[
& \frac{d}{ds} R^\perp=\frac{2 P\bphi^3}{3}  \left( R^\perp-\frac{1}{3} I^\perp  \right) \left( R^\perp+\frac{2}{3} I^\perp  \right), 
\label{eq:renr}\\
& \frac{d}{ds} P^\perp=P\bphi^3 \left[R^\perp P^\perp \right].
\label{eq:renp}
\]

\begin{figure}
\begin{center}
\includegraphics[width=7cm]{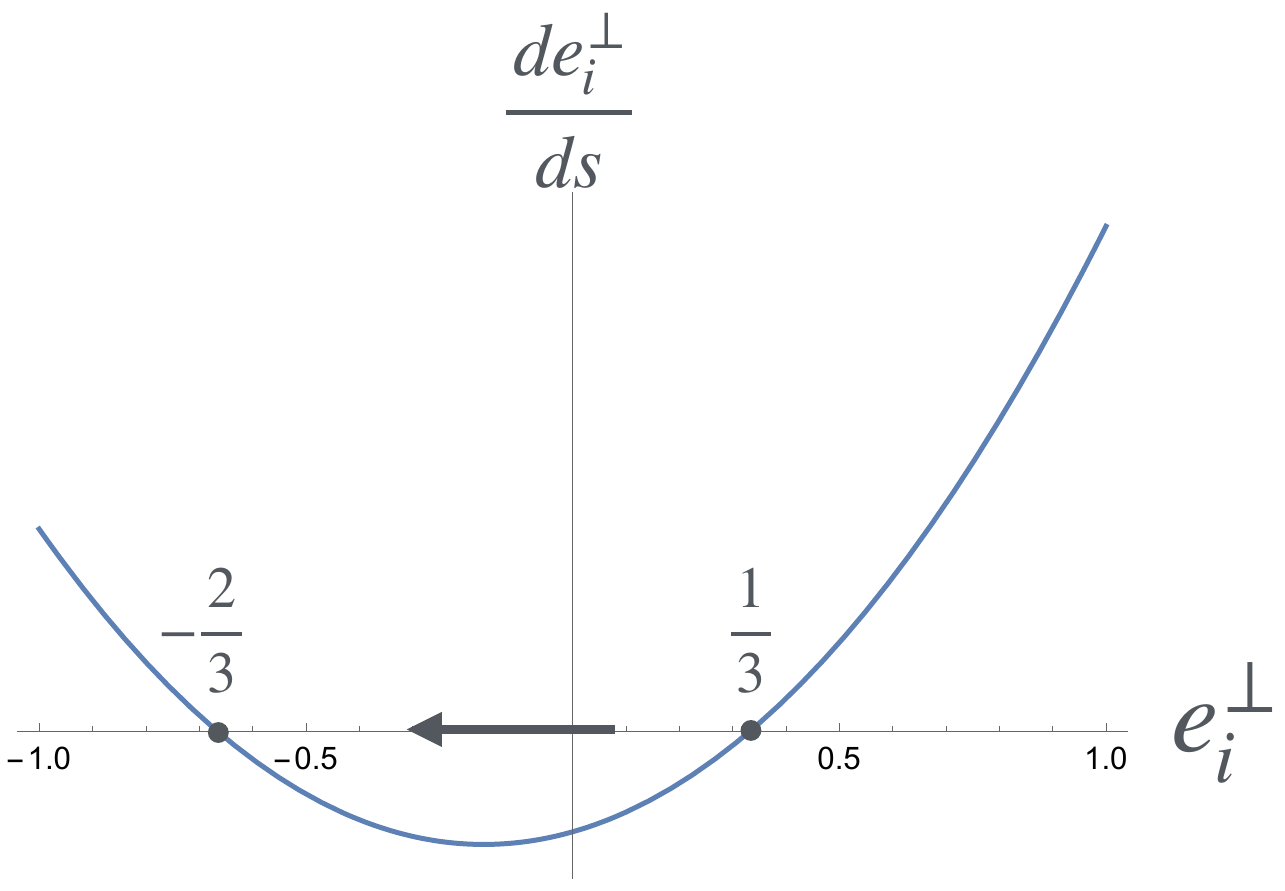}
\caption{The flow of the eigenvalues of $R^\perp$. There is an ultraviolet fixed point at $\frac{1}{3}$, and an infrared one at 
$-\frac{2}{3}$.}
\label{fig:flowR}
\end{center}
\end{figure}
Since $R^\perp$ can be diagonalized in terms of the eigenvalue subspaces, 
the flow equation \eq{eq:renr} can be regarded as the flow equation for the eigenvalues. 
By recalling the convention \eq{eq:convphi} and regarding $s$ as a renormalisation parameter which increases in the 
infrared direction, \eq{eq:renr} implies that $e^\perp_i$ have the ultraviolet fixed point at the eigenvalue $\frac{1}{3}$ and 
the infrared one at $-\frac{2}{3}$ (See Fig.~\ref{fig:flowR}). Note that, because of \eq{eq:eineq}, the eigenvalues exist exactly in the 
middle region where the right-hand side of \eq{eq:renr} is negative or zero, and the flow equation proves that 
the bound \eq{eq:eineq} is indeed tight.
Since the critical points are characterized by the eigenvalue $\frac{1}{3}$ of $R^\perp$ (see \eq{eq:criticaleg}), 
the flow equation \eq{eq:renr} implies that the critical points appear as the ultraviolet fixed points of the flow.  
Thus we have 
\begin{itemize}
\item
The fixed points of the flow can be labeled by $(N_+,N_{-})$ satisfying $N_++N_-=N-1$,
 where $N_+$ and $N_-$ denote the degeneracy of the eigenvalue $\frac{1}{3}$ of $R^\perp$ 
 and that of $-\frac{2}{3}$, respectively.
Starting from an infinitesimal neighborhood of a fixed point, the flow asymptotically goes to a new fixed point 
with smaller $N_+$ (and larger $N_-$).
\end{itemize}

Let us now turn to the flow equation of $P^\perp$ in \eq{eq:renp}. First of all note that, from the above discussions on $R^\perp(s)$,
the decomposition $V=V_\parallel\oplus V_+ \oplus V_- \oplus V_\pp$ discussed in Section~\ref{sec:decompP} 
does not change under the flow. This means that the form of the decomposition of $P$, \eq{eq:pdecomp} and \eq{eq:pperpgen}, 
does not change, though the values may change.  For each of $P^{ijk}\ (i,j,k=+,-,\pp)$ in  \eq{eq:pperpgen}, 
\eq{eq:renp} implies
\[
\frac{d}{ds} P^{ijk}(s) = \frac{P\bphi^3}{3}\left(e^i(s) +e^j(s)+ e^k(s) \right)P^{ijk}(s),
\label{eq:dpds}
\]
where the tensor indices are suppressed for notational simplicity, 
and we have explicitly written the dependence on $s$. Here
the eigenvalues can take values, $e^+=1/3,e^-=-2/3,-2/3<e^\pp<1/3$. Note that 
the notation is rather abusive in the sense that $e^i\neq e^j$ can happen, even when $i=j=\pp$, if $e^i$ and $e^j$ 
correspond to different eigenvalues of $R^\pp$ (or different tensor indices).
But we will use this convenient notation with suppressed tensor indices, because it causes no confusions in the following discussions.

The solution to \eq{eq:dpds} is given by
\[
P^{ijk}(s)=P^{ijk}(s_0) \exp \left(\frac{P\bphi^3}{3} \int_{s_0}^s dt \left( e^i(t) +e^j(t)+ e^k(t) \right) \right),
\label{eq:psols}
\]
where $s_0$ is an arbitrary initial point, and the $s$-dependence of the eigenvalues is given by
\[
e^i(s)=\frac{1}{3}\cdot \frac{ -2 \left(\frac{1}{3}-e^i_0 \right)+\left(\frac{2}{3}+e^i_0\right) e^{-\frac{2 P\bphi^3}{3}(s-s_0)} }{\frac{1}{3}-e^i_0+\left( \frac{2}{3} +e^i_0 \right)e^{-\frac{2 P\bphi^3}{3}(s-s_0)} },
\label{eq:egexp}
\]
which can be obtained by explicitly solving \eq{eq:renr} with $e_0^i=e^i(s_0)$.
One can indeed check the constancy of $e^+(s)=1/3$ and $e^-(s)=-2/3$, 
and $e^\pp(s)\rightarrow -2/3 \hbox{ for }
s\rightarrow \infty$. 

Let us first discuss the $s\rightarrow \infty$ limit. As discussed in Section~\ref{sec:decompP}, $P^\perp$ has generally 
the form \eq{eq:pperpgen}. As for $P^{ijk}=P^{+-\pp},P^{+\pp\pp},P^{-\pp\pp},P^{\pp\pp\pp}$,
there exist $c_1,s_1$ such that $e^i(s)+e^j(s)+e^k(s)<c_1 <0  \hbox{ for } {}^\forall s>s_1$, 
because $e^\pp(s) \rightarrow -2/3$.
Therefore the exponent on the right-hand side of \eq{eq:psols} diverges to $-\infty$ for $s\rightarrow \infty$, and we therefore obtain
\[
P^{+-\pp}(s),P^{+\pp\pp}(s),P^{-\pp\pp}(s),P^{\pp\pp\pp}(s) \rightarrow 0\hbox{ for }s\rightarrow \infty.
\]
 
On the other hand, $P^{++-}(s)$ is constant, because $2 e^++e^-=0$. As for $P^{++\pp}$, we have
\[
2 e^++e^\pp(s)=\frac{\left(\frac{2}{3}+e^\pp_0\right)e^{-\frac{2 P\bphi^3}{3}(s-s_0)} }{
\left(\frac{1}{3}-e^\pp_0\right)+\left(\frac{2}{3}+e^\pp_0\right)e^{-\frac{2 P\bphi^3}{3}(s-s_0)}}.
%<\frac{\left(\frac{2}{3}+e^\pp_0\right)}{\left(\frac{1}{3}-e^\pp_0\right)}e^{-\frac{2 P\bphi^3}{3}(s-s_0)}. 
\]
Since the exponent in \eq{eq:psols} converges in $s\rightarrow\infty$ in this case, we obtain $P^{++\pp}(s)\rightarrow \hbox{finite}$. 
Therefore, considering $e^\pp(s)\rightarrow -2/3=e^-$, $P^{++\pp}(s)$ asymptotically joins $P^{++-}$ in the $s\rightarrow \infty$ limit. Summarizing all the results above, 
we conclude $P^\perp(s) \rightarrow P^{++-} \hbox{ for }s\rightarrow \infty$.

Let us next discuss the opposite limit $s\rightarrow -\infty$. In this case $e^\pp(s)\rightarrow e^+=1/3$.
The exponent in \eq{eq:psols} negatively diverge, if there exist real numbers $c_1, s_1$ such that $e^i(s) +e^j(s)+e^k(s)>c_1>0$ for 
$^\forall s<s_1$. Therefore, with similar discussions as above, we conclude $P^{++\pp}(s),P^{+\pp\pp}(s),P^{\pp\pp\pp}(s) 
\rightarrow 0$,
while $P^{+-\pp}(s),P^{-\pp\pp}(s)\rightarrow \hbox{ finite}$, asymptotically joining $P^{++-}$ in the limit. 
Therefore we find $P^\perp(s) \rightarrow P^{++-} \hbox{ for }s\rightarrow -\infty$.

The discussions above and some former statements conclude
\begin{itemize}
\item
Every flow line is asymptotically terminated by fixed points at both ends.  Fixed points are characterized by a pair of integers 
$(N_+,N_-)$ with $N_++N_-=N-1$. When the upstream and downstream fixed points of a flow line 
have $(N_+,N_-)$ and $(N_+',N_-')$, respectively, they satisfy $N_+>N_+'$ (and $N_-<N_-'$).
\item
A fixed point with $(N_+,N_-)$ has the decomposition,
\s[
P=\frac{8 P\bphi^3}{27} \bphi\otimes \bphi \otimes \bphi + \frac{2 P\bphi^3}{3}[ \bphi\otimes I^+]- \frac{4 P\bphi^3}{3}[ \bphi\otimes I^-]+
P^{++-}.
\label{eq:fixedptp}
\s]
Here the tensor index space is decomposed as $V=V_\parallel\oplus V_+ \oplus V_-$, where the dimensions of $V_+$ and $V_-$
are $N_+$ and $N_-$, respectively. $I^+$ and $I^-$ are the projections onto $V_+$ and $V_-$, respectively. 
$P^{++-} \in [V_+\otimes V_+ \otimes V_-]$ and therefore its dimension is $N_+ (N_++1) N_-/2$. 
\end{itemize}

For $N_+,N_->0$ $P^{++-}$ has a finite dimension. The components can freely be taken, unless a bound is violated.
To obtain the bound let us consider an arbitrary vector $\tphi$ of size $|\tphi|^2=3/2$, which can be
parameterized as $\tphi=\bphi \cos \theta+\eta_+ |\bphi| \sin\theta \cos \varphi+\eta_- |\bphi| \sin\theta \sin \varphi$, where
$\eta_+\in V_+,\eta_-\in V_-$ with $|\eta_+|=|\eta_-|=1$. For \eq{eq:fixedptp}, we obtain
\[
P\tphi^3=P\bphi^3\left( \cos^3\theta+\frac{3}{2} \cos \theta \sin^2\theta \cos^2 \varphi-3 \cos \theta \sin^2\theta \sin^2 \varphi +\frac{3 |\bphi|^3 P^{++-} \eta_+^2 \eta_- }{P\bphi^3}\sin^3 \theta \cos^2 \varphi \sin \varphi \right).
\label{eq:ppt3}
\]
The inequality, $P\tphi^3 \leq P\bphi^3$, must hold for all $\theta,\varphi, \eta_+,\eta_-$ because of \eq{eq:newfree}. By putting 
$\theta=\pi/2,\varphi=\arccos(\sqrt{2/3})$, we obtain a necessary condition,
\[
\max_{|\eta_+|=|\eta_-|=1} P^{++-} \eta_+^2 \eta_ - \leq \frac{\sqrt{3} P\bphi^3}{2 |\bphi|^3}.
\label{eq:maxppm}
\]
In fact, as proven in \ref{app:maxppm}, this is also sufficient. 

The form \eq{eq:fixedptp} can be used to show some connections between the fixed points and the first-order phase transition surfaces.
To avoid an exceptional case let us assume that the inequality \eq{eq:maxppm} is not saturated, namely,
\[
\max_{|\eta_+|=|\eta_-|=1} P^{++-} \eta_+^2 \eta_ - < \frac{\sqrt{3} P\bphi^3}{2 |\bphi|^3}.
\label{eq:strict}
\]
Then we can prove
\begin{itemize}
\item
The fixed points with $N_->0$ are on the first-order phase transition surfaces.
\item
The critical fixed points with $N_+>0$ are on the edges of the first-order phase transition surfaces.   
\end{itemize}
The first statement above can be proven by using a result from \ref{app:maxppm}. For \eq{eq:strict} there exist 
three distinctly located maxima of $P\phi^3$ (or $h$), which are the first two solutions with $p=1$ in \eq{eq:normxyz}.  
Therefore the fixed point is where three different phases coexist.
It is also possible to see the phase transitions explicitly by unbalancing their values of $P\phi^3$ by perturbations: 
The original state $\bphi$ becomes the unique maximum by adding $\bphi\otimes \bphi \otimes \bphi$ to $P$ with a positive coefficient, 
and the other two phases by adding $P^{---}$ with either sign. 
Therefore the fixed point is a meeting point of the first-order phase transition surfaces.
The proof of the second statement above is given in \ref{app:first}.

In the discussions of \ref{app:maxppm} we notice that the following property holds, if \eq{eq:maxppm} is saturated:
\begin{itemize}
\item
If $\max_{|\eta_+|=|\eta_-|=1} P^{++-} \eta_+^2 \eta_ - =\frac{\sqrt{3} P\bphi^3}{2 |\bphi|^3}$,
$\bphi$ has a continuous locus. 
\end{itemize}

\section{Critical exponents}
\label{sec:criticalexp}
In this section we will study the scaling properties of the perturbations around the fixed points. 

We first have to remove the gauge redundancy to determine the scalings unambiguously. 
The gauge transformations are the orthogonal group transformations \eq{eq:ontrans} in the index vector space, and 
its dimension is given by $N(N-1)/2$. 
The form of $P$ given in \eq{eq:genP1} has already taken into account some gauge degrees of freedom by 
representing it by using $\bphi$, consequently making $v^\perp$ vanish as proven in \eq{eq:avp}, where $N-1$ gauge degrees of freedom have
been consumed. 
The remaining gauge degrees of freedom can be used to diagonalize $R^\perp$, making the off-diagonal components vanish. 
In fact, the number of the vanished components agrees with the dimension of the gauge degrees of freedom: $N-1+(N-1)(N-2)/2=N(N-1)/2$.

Let us now consider a $P$ which is close to a fixed point with $(N_+,N_-)$, where $N_++N_-=N-1$. 
From the form of \eq{eq:genP1} and \eq{eq:fixedptp} at the fixed point, 
and by ordering the eigenvalues of $R^\perp$ in a convenient manner, $P$ can be parameterized with no gauge redundancy by 
\s[
&e^+_{a_+} =\frac{1}{3}-\delta e^+_{a_+} , \ a_+=1,2,\cdots,N_+, \\
&e^-_{a_-}=-\frac{2}{3}+\delta e^-_{a_-} , \ a_-=1,2,\cdots,N_-, \\
&P^\perp=P^{++-}_0+\delta P^{+++}+\delta P^{++-}+\delta P^{+--}+\delta P^{---},
\label{eq:paraperturb}
\s]
where $e^+_{a_+}$ and $e^-_{a_-}$ are the eigenvalues of $R^\perp$ which take values close to the fixed point values, $1/3$ and $-2/3$,
respectively, the variables with $\delta$ are small perturbations, and $P^{++-}_0$ is the value of $P^{++-}$ at the fixed point. 
Here we have introduced $a_+$ and $a_-$ for the indices in $V_+$ and $V_-$, respectively. 
By applying \eq{eq:renr} and \eq{eq:renp} (or \eq{eq:dpds}) and taking the lowest order of the perturbations, 
we obtain
\s[
&\frac{d}{ds}\delta e^+=\frac{2 P\bphi^3}{3} \delta e^+, \\
&\frac{d}{ds} \delta e^-=-\frac{2 P\bphi^3}{3} \delta e^-, \\
&\frac{d}{ds}\delta  P^{+++}=\frac{ P\bphi^3}{3} \delta P^{+++}, \\
&\frac{d}{ds}\delta  P^{+--}=-\frac{ P\bphi^3}{3} \delta P^{+--}, \\
&\frac{d}{ds}\delta  P^{---}=-\frac{ 2P\bphi^3}{3} \delta P^{---},
\label{eq:dsep}
\s]
where the vector space indices are suppressed, because the developments are independent of the indices. 
This shows that $\delta e^+$ and $\delta P^{+++}$ are relevant, while the others above are 
irrelevant. Therefore the dimension of the relevant and irrelevant perturbations are given by
\s[
&D_{\rm relevant}=\frac{N_+ ( N_+^2+ 3 N_+ +8)}{6}, \\
&D_{\rm irrelevant}=\frac{N_- (8 + 3 N_- + N_-^2 + 3 N_+ + 3 N_+ N_- )}{6},
\label{eq:Drelev}
\s]
respectively. The ratio of the larger and the smaller scalings of the relevant directions give the critical exponent,
\[
\nu_c=\frac{1}{2},
\]
which is indeed what is expected from the mean field anaysis.

As for $P^{++-}$ we need more careful analysis, considering dependence on indices.
From \eq{eq:dpds} and \eq{eq:paraperturb}, 
we obtain in the lowest order 
\[
\frac{d}{ds} \delta P^{++-}_{a_+b_+c_-} =\frac{2 P \bphi^3}{3} (-\delta e^+_{a_+}-\delta e^+_{b_+}+\delta e^-_{c_-}) P^{++-}_{0\,a_+b_+c_-}.
\label{eq:dspppm}
\]
This seems to show that $\delta P^{++-}$ 
develops. However, there is the possibility of reparameterization. Let us introduce 
\[
\delta \tilde P^{++-}_{a_+b_+c_-}\equiv \delta P^{++-}_{a_+b_+c_-}+(\delta e^+_{a_+}+\delta e^+_{b_+}+\delta e^-_{c_-}) 
P^{++-}_{0\, a_+b_+c_-},
\]
which is a new independent variable replacing $\delta P^{++-}$.
Then from \eq{eq:dsep} and \eq{eq:dspppm} we obtain 
\[
\frac{d}{ds} \delta \tilde P^{++-}_{a_+b_+c_-}=0.
\]
Therefore $\delta \tilde P^{++-}$ are marginal and its dimension  is given by
\[
D_{\rm marginal}=\frac{N_-  N_+(N_++1)}{2}.
\]

\section{An analog of RG-decreasing function}
\label{sec:analog}
Let us define a function,
\[
d_{\rm RG}={\rm Tr} \left( R+\frac{2}{3} I \right)-\frac{4}{3}={\rm Tr} \left( R^\perp+\frac{2}{3} I^\perp \right).
\label{eq:defofC}
\]
Let us consider a flow line parameterized by $s$ as in the former sections. 
Then, since $R$ has the eigenvalue $2/3$ in the parallel direction, and some of the others monotonically decrease 
with $s$ (See Section~\ref{sec:flow}), 
$d_{\rm RG}(s)$ is a monotonically decreasing function of $s$ on the flow line.

On a fixed point with label $(N_+,N_-)$,  $R^\perp$ has the eigenvalues $1/3$ and $-2/3$ with degeneracy $N_+$ and $N_-$, respectively.
Therefore
\[
d_{\rm RG}^{\rm fixed\, pt.}=N_+
\label{eq:cfixed}
\] 
on the fixed point. In general,
\[
0\leq d_{\rm RG} \leq N-1,
\]
and the bound is tight: the minimum and the maximum are realized on the fixed points with $(N_+,N_-)=(0,N-1)$ and $(N-1,0)$,
respectively. 

Since $R$ depends on $\bphi$ as defined in \eq{eq:defofR},  $d_{\rm RG}$ is not generally unique
on the first-order phase transition surfaces. This multiplicity 
does not ruin the monotonic decrease of $d_{\rm RG}(s)$ along the flow lines, 
since they cannot cross the first-order phase transition surfaces, as proven in Section~\ref{sec:firstorder}. 
 
A field theoretical interpretation of \eq{eq:cfixed} is that $d_{\rm RG}^{\rm fixed\, pt.}$ counts the number of 
massless modes, since $N_+$ is the degeneracy of the zero eigenvalue of the Hessian matrix $K$ as in \eq{eq:criticaleg}.
The decrease of $d_{\rm RG}(s)$ along a flow line can be understood as the process that some of 
the modes become more and more massive, and 
asymptotically disappears from $d_{\rm RG}(s)$ in $s\rightarrow \infty$. 

The multiplicity of $d_{\rm RG}$ on the first-order phase transition surfaces can naturally be understood field theoretically, 
because spectra of modes generally depend on phases. 

\section{Ambiguity of $(N_+,N_-)$ and cyclic flows}
\label{sec:deg}
As defined in \eq{eq:defofR},  the eigenvalues of $R^\perp$ depend on $\bphi$. Therefore, on the first-order phase transition surfaces, 
the label $(N_+,N_-)$ may not be unique, and it is even possible that a fixed point in one phase  
may not be so in another, since $R^\perp$ may have eigenvalues different from the fixed point values, 
$-2/3$ or $1/3$, in the latter phase.  In this section we will study these matters. 

Let us consider a fixed point labeled by $(N_+,N_-)$ with $N_++N_-=N-1$, and assume $N_->0$ and \eq{eq:strict},
namely, the bound is not saturated.
Then, as discussed in Section~\ref{sec:flow}, the fixed point is located on a first-order phase
transition surface. From \eq{eq:normxyz} three cases coexist there and respectively have the ground states, 
$\bphi$ and $\tphi= -\bphi/2+ \frac{\sqrt{3}|\bphi|}{2} \eta_-\ (\eta_-\in V_-,\ |\eta_-|=1)$, where the $\pm$ sign 
in the second case of \eq{eq:normxyz} has been absorbed into $\eta_-$. 
Since $|\tphi|=|\bphi|$,  $\tphi$ can be obtained from $\bphi$ by a rotation on the $(\bphi,\eta_-)$ plane. 
To describe the rotation more explicitly
let us introduce $\tilde \eta_-=-\sqrt{3} \bphi/(2|\bphi|)-\eta_-/2$, which corresponds to the rotated $\eta_-$. Then 
\s[
\bphi=-\frac{1}{2} \tphi -\frac{\sqrt{3}}{2} |\tphi| \tilde \eta,\ 
\eta_-=\frac{\sqrt{3}}{2} \frac{\tilde \phi}{|\tphi|} -\frac{1}{2} \tilde \eta.
\label{eq:btphi}
\s]
Note that $\tphi$ and $\tilde \eta$ are transverse to each other, and also $P\tphi^3=P\bphi^3$, 
because the three cases have the same value of $P\phi^3$.
Putting \eq{eq:btphi} into \eq{eq:fixedptp}, we obtain
\s[
P=&\frac{8 P \tphi^3}{27} \tphi \otimes \tphi \otimes \tphi+ \frac{2 P\tphi^3}{3}[ \tphi\otimes I^{-\perp \eta_-}]
-\frac{4 P\tphi^3}{3} [ \tphi \otimes \teta \otimes \teta ] -\frac{P\tphi^3}{3} [\tphi \otimes I^+] \\
&
+\frac{3}{\sqrt{2}} [ \tphi\otimes (P^{++-}\eta_-) ]+\frac{2P\tphi^3 |\tphi|}{\sqrt{3} } [\teta\otimes I^{-\perp\eta_-} ] 
-\frac{P\tphi^3 |\tphi | }{\sqrt{3}} [\teta \otimes I^+]  -\frac{3}{2} [\teta\otimes (P^{++-} \eta_-) ] \\
&+3 [P^{++-} I^{-\perp \eta_-}]. 
\label{eq:ptilde}
\s]
To get this expression we have decomposed $I^-$ into the projections onto $\eta_-$ and the transverse subspace:
\[
I^-=\eta_-\otimes \eta_- + I^{-\perp \eta_-}.
\label{eq:decompmin}
\]
We have also used 
a formula\footnote{This formula can be derived as follows. $P^{++-}$ is a symmetric tensor defined by
$P^{++-}_{abc}=I^+_{aa'}I^+_{bb'} I^-_{cc'} P_{a'b'c'}+
I^+_{aa'}I^-_{bb'} I^+_{cc'} P_{a'b'c'}+I^-_{aa'}I^+_{bb'} I^+_{cc'} P_{a'b'c'}$. 
Therefore $[P^{++-}]=3 [P^{++-} I^-]$. Putting \eq{eq:decompmin} into this, we obtain the formula.},
\[
[P^{++-}]=3[P^{++-} I^-]=3 [(P^{++-}\eta_- )\otimes \eta_-]+3 [P^{++-}I^{-\perp \eta_-}],
\]
where $(P^{++-}\eta_-)_{ab}=P^{++-}_{abc}\eta_{-c}$, $(P^{++-}I^{-})_{abc}=P^{++-}_{abc'} I^{-}_{c'c}$, and similarly  for 
$(P^{++-}I^{-\perp \eta_-})_{abc}$.
In the expression \eq{eq:ptilde}  $P$ is represented according to the orthogonal decomposition $V=V_{\tilde \phi}\oplus V_+ \oplus V_{\tilde \eta} \oplus
V_{-\perp \eta_-}$, where $V_{\tphi}$ and $V_{\tilde \eta}$ are the one-dimensional subspaces 
along the $\tphi$ and $\tilde \eta$ directions, respectively,
and $V_{-\perp \eta_-}$ is the subspace of $V_-$ transverse to $\eta_-$. 

We are now interested in the spectra of $R^\perp$ for $\tphi$. From the definition \eq{eq:defofR}, \eq{eq:Rdecomp} and \eq{eq:ptilde}, 
we obtain
\[
\tilde R^\perp=\frac{P \tilde \phi}{P\tilde \phi^3}-\frac{4}{9} \tilde\phi \otimes \tilde \phi=
\frac{1}{3} I^{-\perp \eta}-\frac{2}{3} \teta \otimes \teta +\left(-\frac{1}{6} I^++\frac{3}{2 \sqrt{2}} \frac{P^{++-} \eta_-}{P\bar \phi^3}
\right). 
\label{eq:rtilde}
\]
From the first term we see that $V_{-\perp \eta_-}$ is the subspace of the eigenvalue $1/3$ of $\tilde R^\perp$,
and therefore can be denoted as $\tilde V_+$ for $\tphi$ with the same meaning as $V_+$ for $\bphi$. 
Its dimension is $\dim \tilde V_+=\dim V_{-\perp \eta_-}=N_--1$. 
From the second term we see that $V_{\tilde \eta}$ is the subspace of the eigenvalue $-2/3$, and 
serves as $\tilde V_-$, whose dimension is $\dim \tilde V_-=\dim V_{\teta}=1$. As for the third term, it is a matrix in $V_+$. 
The condition \eq{eq:strict} implies that the absolute values of the eigenvalues of $P^{++-}  \eta_ -$ are bounded by the right-hand side
of \eq{eq:strict}. 
Therefore the eigenvalues of the third term are bounded within the region $(-2/3,1/3)$, where
the boundaries are not included. This means that $V_+$ serves as $\tilde V_\pp$, whose dimension is 
$\dim \tilde V_\pp= \dim V_+=  N_+$. 
Therefore, when $N_+>0$, $P$ is not a fixed point in the phase with the ground state $\tphi$, while it is so in the phase with $\bphi$.

For convenience of the following discussion let us use the same label $(n_+,n_-)$ based on the degeneracies of the eigenvalues 
$1/3$ and $-2/3$ of $R^\perp$ for a non-fixed point as well. So we allow $n_++n_-<N-1$. 
Then $P$ can be labeled as $(N_--1,1)$ in the phase with $\tphi$ according to $\dim \tilde V_+$ and $\dim \tilde V_-$
obtained above.
In the phase with $\tphi$, starting from $P$, the flow asymptotically approaches a fixed point with label $(N_--1,N-N_-)$.
Thus, by allowing a transition of $P$ near a fixed point from the phase $\bphi$ to $\tphi$, 
which can generally be realized by an infinitesimal jump of $P$ near a fixed point, 
the following flow from an infinitesimal vicinity of a fixed point to another can be constructed:
\[
(N_+,N_-) \Rightarrow (N_--1,1) \longrightarrow(N_--1,N-N_-),
\label{eq:singleflow}
\]
where the infinitesimal jump between phases and the continuous flow in one phase 
are denoted by  $\Rightarrow$ and $\rightarrow$, respectively.
More generally, since the degeneracy of the eigenvalues $1/3$ and $-2/3$ of $R^\perp$ can easily be deducted by an infinitesimal change of 
$P$\footnote{For example, one may add an infinitesimal tensor, $-\sum_{i=1}^{N_+} 
\epsilon^+_i [\bphi \otimes \eta^+_i \otimes \eta^+_i]
+\sum_{i=1}^{N_-} 
\epsilon^-_i [\bphi \otimes \eta^-_i \otimes \eta^-_i]\ (\epsilon^+,\epsilon^-\geq 0, \eta^+_i \in V_+,  \eta^-_i \in V_-)$ 
to \eq{eq:pdecomp}.},
we can also construct
\[
(N_+,N_-) \Rightarrow (\tilde N_+,\tilde N_-) \longrightarrow(\tilde N_+ ,N-\tilde N_+ -1),
\label{eq:singleflowgen}
\]
where $\tilde N_+ \leq N_--1$ and $\tilde N_-\leq 1$.

If $N_+ \leq \tilde N_+$ the endpoints of the flow in \eq{eq:singleflowgen} cannot be realized by a continuous flow in one phase,
as discussed in Section~\ref{sec:flow}.
In particular, if $N_+\leq N_--1$, one can take $\tilde N_+=N_+$, and construct a cyclic flow,
\[ 
(N_+,N_-) \Rightarrow (N_+,\tilde N_- )  \longrightarrow(N_+ ,N_-).
\label{eq:particularcyclic}
\]
We will show an explicit example of a cyclic flow in Section~\ref{sec:cyclicex}.

The above flow \eq{eq:singleflowgen} does not contradict the monotonic decrease of the RG-function $d_{\rm RG}(s)$. 
This is because it also depends on $\bphi$ and is therefore multi-valued on the first-order phase transition surfaces. 
The value can have a discrete jump under an infinitesimal jump of $P$ near the first-order phase transition surfaces. 

\section{The flow equation as the $n\rightarrow \infty$ limit of an identity}
\label{sec:identity}
An essential property of a renormalization group flow is the invariance of a theory under the simultaneous change of renormalization
scale and couplings. This suggests that the flow equation introduced in Section~\ref{sec:floweq} can actually be derived 
from an identity. In this section, we will prove an identity in our system,
and its $n\rightarrow \infty$ limit indeed derives the flow equation. 

For convenience, let us introduce a derivative operator $D$ with respect to $P$ as
\[
D_{abc} P_{def}=\{ M_{abc},P_{def} \} =\frac{1}{6}\sum_{\sigma} \delta_{a\,\sigma_d}\delta_{b\,\sigma_e}\delta_{c\,\sigma_f},
\label{eq:defofder}
\] 
where the sum over $\sigma$ is over all the permutations of $d,e,f$.
Let us apply the first term of $\phi_a {\cal H}_a$, namely, $\phi_a P_{abc}P_{bde} M_{cde}$ in \eq{eq:hamiltonian},
to the expression \eq{eq:rescaleZ}. 
By performing a partial integration of $\phi$, one obtains
\s[
\int_{P\phi^3>0}  d^N &\phi \  \left\{ \phi_a P_{abc}P_{bde} M_{cde} , e^{-n(\phi^2-\log(P\phi^3))} \right\}\\
&=\int_{P\phi^3>0} d^N \phi \ \phi_a P_{abc}P_{bde} D_{cde} \, e^{-n(\phi^2-\log(P\phi^3))} \\
&=\int_{P\phi^3>0}  d^N\phi\ n \,\phi_a P_{abc}P_{bde}\frac{\phi_c \phi_d \phi_e}{P\phi^3}  e^{-n(\phi^2-\log(P\phi^3))} \\
&=\int_{P\phi^3>0}  d^N \phi\ \frac{n(P\phi^2)_a (P\phi^2)_a}{P\phi^3}  e^{-n(\phi^2-\log(P\phi^3))}\\
&=\int_{P\phi^3>0}  d^N \phi\ \frac{1}{3} (P\phi^2)_a  e^{-n\phi^2}\partial_a e^{n\log(P\phi^3)} \\
&=\frac{2}{3} \int_{P\phi^3>0}  d^N \phi\ \left( -(P\phi)_{aa}+n (P\phi^3)\right) e^{-n(\phi^2-\log(P\phi^3))}.
\label{eq:applyfirst}
\s]

A comment is in order. In the derivation of \eq{eq:applyfirst} we have assumed that the boundary contributions can be ignored 
in performing partial integrations. This can be justified at the boundary $P\phi^3=0$, because the integrand
contains $(P\phi^3)^n$. The other boundary contribution from $\phi\rightarrow\infty$ can also be ignored 
because of $e^{-n \phi^2}$ in the integrand.

Let us next consider the cosmological term in \eq{eq:hamiltonian}. Let us allow $\lambda$
to be dependent on $P,\phi$ as $\lambda=\lambda(P,\phi)$. Then, 
\[
\int_{P\phi^3>0}  d^N \phi\, \left\{ \lambda(P,\phi) \phi_a M_{abb}, e^{-n(\phi^2-\log(P\phi^3))} \right\}
&= \int_{P\phi^3>0}  d^N \phi\,  \lambda(P,\phi) \phi_a D_{abb} e^{-n(\phi^2-\log(P\phi^3))} \nonumber \\
&=\int_{P\phi^3>0}  d^N \phi\,   \lambda(P,\phi) \frac{n (\phi^2)^2}{P\phi^3} e^{-n(\phi^2-\log(P\phi^3))}.
\label{eq:applycos}
\]
Then, by choosing 
\[
\lambda(P,\phi)=\frac{2(P\phi^3)^2}{3 (\phi^2)^2},
\label{eq:lampphi}
\]
the second term of \eq{eq:applyfirst} can be canceled by subtracting \eq{eq:applycos}.

As for the first term of \eq{eq:applyfirst}, it can simply be 
canceled by adding $\frac{2}{3} P_{abb}$ to the Hamiltonian operator. Thus, we obtain 
an identity,
\[
\int d^N\phi\,  \phi_a \hat{\cal H}_a e^{-n(\phi^2-\log(P\phi^3))}=0
\label{eq:exactRG}
\]
with
\[
\hat {\cal H}_a=P_{abc}P_{bde}D_{cde}+\frac{2}{3} P_{abb} -\lambda(P,\phi) D_{abb}.
\label{eq:tildeh}
\]

Now, let us discuss the thermodynamic limit $n\rightarrow \infty$ of \eq{eq:exactRG}.
The consequence of the thermodynamic limit is that the integration can be replaced by the integrand at
$\phi=\bar \phi$ which maximizes the exponent. Then, by using 
\eq{eq:phi2}, \eq{eq:lamval} and \eq{eq:lampphi}, we find 
 \[
\lambda(P,\bar\phi)=\frac{8 (P\bar \phi^3)^2}{27}=\bar \lambda.
\] 
In addition the second term of \eq{eq:tildeh} can be ignored in the $n\rightarrow \infty$ limit, being compared with the
other terms. This is because the others contain $D_{abc}$, which induces a factor of $n$ by taking
derivatives of the exponent, while the second term does not.  
Then \eq{eq:tildeh} can be identified with \eq{eq:hamiltonian}.
From these considerations, in $n\rightarrow \infty$, \eq{eq:exactRG} implies
\[
\frac{d}{ds}Z_n(P)=
\left\{\bar \phi_a {\cal H}_a, Z_n(P)\right\} =Z_n\left(\left\{\bar \phi_a {\cal H}_a, P\right\}\right)=0
\]
with $\lambda=\bar \lambda$. 
Thus, the flow \eq{eq:Pflow} is an invariant flow of the partition function of RCTN in the 
thermodynamic limit, which would entitle the flow as an RG flow.

%%%%%%%%%%%%%%%%%%%%%%%%%%%%%%%%%%%%%%%%%%

\section{Examples}
\label{sec:examples}
In this section we will explicitly study the examples of $N=2,3$. 
As discussed in Section~\ref{sec:criticalexp}, it is important to remove gauge redundancy to unambiguously study the 
physical properties. Let us first describe how we draw the flow with no gauge redundancy.

\subsection{Flow equation with gauge-fixing}
\label{sec:gfixflow}
From \eq{eq:hamiltonian} and \eq{eq:lamval}, the Hamiltonian we are considering has the form,
\[
\bar H=\bar \phi_a P_{abc}P_{bde}M_{cde}- \frac{8}{27} (P \bar\phi^3)^2 M_{abb}.
\] 
Let us introduce
\[
{\cal D}= P_{abc}M_{abc},
\]
which is a generator of a scale transformation, $\{{\cal D},P_{abc}\}=P_{abc}$ and $\{{\cal D},M_{abc}\}=-M_{abc}$ \cite{Sasakura:2013gxg}.
Since \eq{eq:stationary} shows that $\bar \phi$ does not depend on the overall scale of $P$, 
we obtain $\{ {\cal D}, \bar \phi\}=0$. So we get
\[
\{ {\cal D},\bar H\}=\bar H.
\label{eq:dhh}
\]

This scale transformation is a physical symmetry of RCTN, because the scale transformation of $P$ merely changes the overall 
factor of the partition function \eq{eq:ZofRCTN}, and therefore does not change its physical properties.
In addition, the $O(N,\mathbb{R})$ transformation \eq{eq:ontrans} is the gauge symmetry, and its Lie generators are given 
by
\[
{\cal J}_{ab}=\frac{1}{2} \left( 
P_{acd} M_{bcd}-P_{bcd} M_{acd} 
\right),
\]
which satisfy $\{ {\cal J}_{ab}, \bar H \}=0$ and $\{ {\cal J}_{ab}, {\cal D} \}=0$.
Since these are the gauge symmetries of our system keeping its physical properties,  
we can consider the following generalization of the flow, 
\[
\begin{split}
\frac{d}{ds}P_{abc}&=\{\tilde H,P_{abc}\}, \\ 
\tilde H&= \bar H + r \, {\cal D}+w_{ab} {\cal J}_{ab},
\end{split}
\label{eq:genflow}
\]
where $r,w_{ab}\,(=-w_{ba})$ are real parameters.
In case a gauge-fixing condition is imposed on $P$, these parameters should be tuned so that $P$ keeps satisfying it along the flow.

The closure of the Poisson algebra among $\bar H, \, {\cal D},\,{\cal J}$ implies the gauge equivalence of the flow equation. 
To see this explicitly, 
let us introduce their operator versions $\hat {\bar H}, \, \hat{\cal D},\,\hat{\cal J}$ which are obtained by replacing $M_{abc}$ with
$D_{abc}$ (introduced in \eq{eq:defofder}) in ${\bar H}, \, {\cal D},\,{\cal J}$, respectively.
From \eq{eq:dhh} and $[ \hat {\cal J}_{ab}, \hat{\bar H} ]=0$, we obtain
$\hat{\bar H} ( r \hat {\cal D}+w_{ab} \hat J_{ab} )=(-r+r \hat {\cal D}+w_{ab} \hat {\cal J}_{ab})\hat{\bar H}$,
where $[\cdot,\cdot]$ denotes a commutator, $[\hat A,\hat B]=\hat A \hat B-\hat B \hat A$.
Then we can find
\[
e^{s \hat{\bar H}} e^{r \hat {\cal D} +w_{ab} \hat J_{ab}}P=e^{r \hat{\cal D} +w_{ab} \hat J_{ab}} e^{s\, e^{-r} \hat{\bar H}} P.
\]
This proves that the flow lines containing gauge-equivalent $P$'s can be mapped to each other by gauge transformations. 

Another important property of $\tilde H$ is that a fixed point of $\tilde H$ is actually that of $\bar H$. The proof is given in \ref{app:fixedpt}.

\subsection{$N=2$ example}
RCTN can be regarded as statistical models on random networks, and in particular 
the $N=2$ case can describe the Ising model on random
networks 
\cite{Bachas:1994qn,dembo,dembo+,Dorogovtsev:2002ix,Dorogovtsev:2008zz,johnston,leone,Sasakura:2014zwa,Sasakura:2014yoa}.
Suppose Ising spins are on vertices, and they are interacting with the nearest neighbors which are linked by edges.  
Then the corresponding tensor is given by \cite{Sasakura:2014zwa,Sasakura:2015xxa}
\[
P_{abc}=\sum_{i=1}^2 R_{ai} R_{bi} R_{ci} e^{h s_i},
\label{eq:isingp}
\]
where $s_1=1,s_2=-1$, $h$ is a magnetic field, and $R$ is a two-by-two matrix satisfying
\[
(R^T R)_{ij}=e^{J s_i s_j},
\]
with $J$ being the nearest neighbor coupling. For $J\geq 0$, one can always find a real matrix $R$.
 
As for gauge-fixing, we have two gauge degrees of freedom for $N=2$, one from ${\cal D}$ and the other from ${\cal J}$.  
The expression \eq{eq:isingp} is not convenient to track a gauge-fixed flow in the way explained in Section~\ref{sec:gfixflow}. 
A more convenient gauge-fixing condition is given by fixing two of the components of $P$ as \cite{Sasakura:2015xxa}
\[
P_{111}=1,\ P_{112}=0.3,\ P_{122}=x_1,\ P_{222}=x_2,
\label{eq:gfix}
\]
where $x_1,x_2$ are the remaining variables. The number $0.3$ is arbitrarily chosen as a non-trivial example of the gauge-fixing.  

\begin{figure}
\begin{center}
\includegraphics[width=8cm]{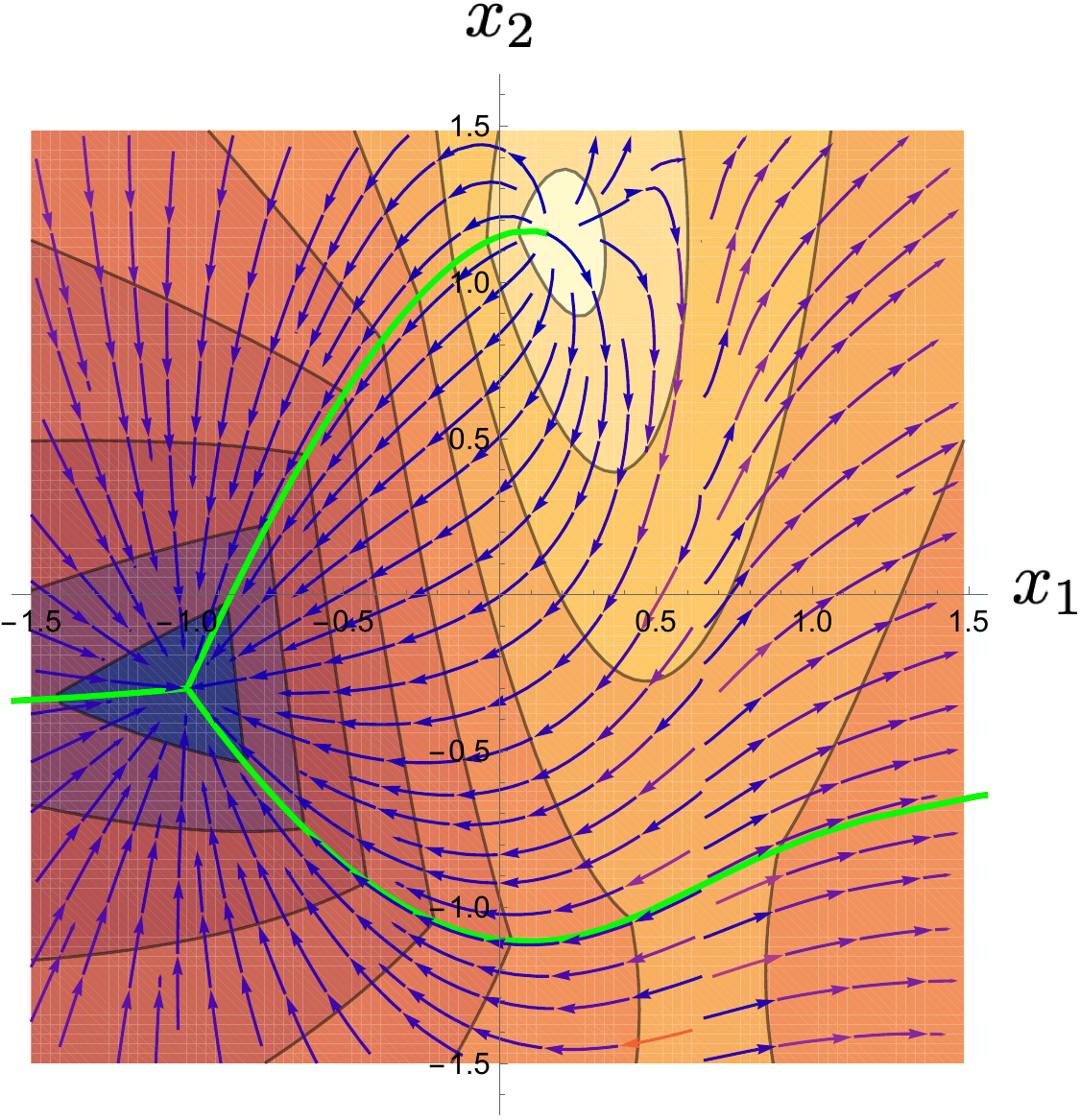}
\caption{The phase structure, the flow and the RG-function in the $N=2$ case.
The solid lines are the first-order phase transition lines. There is a critical fixed point near $(0.2,1.2)$ at the tip of one of the first-order
phase transition lines. 
The RG-function $d_{\rm RG}$ is depicted by a contour plot, and takes one and zero at the upper and the lower fixed points, respectively.
The singular behavior of the flow around $x_1 \sim 0.6$ is due to the singularity of taking the gauge 
\eq{eq:gfix} and has no physical relevance.
}
\label{fig:N=2}
\end{center}
\end{figure}

By a numerical analysis based on Section~\ref{sec:gfixflow} we can find the first-order phase transition lines,  the flow, 
and the RG-function $d_{\rm RG}$, as shown in Figure~\ref{fig:N=2}.
In the figure the tip of a first-order phase transition line near $(x_1,x_2)\sim(0.2,1.2)$ is a critical fixed point, which
is a Curie point of the Ising model. The arrows are depicted by the flow equation \eq{eq:genflow} under the gauge-fixing \eq{eq:gfix}.
The Curie point has the type $(N_+,N_-)=(1,0)$, and it has two dimensional relevant directions, agreeing 
with the formula in \eq{eq:Drelev}.
There is an absorption point of the flow, which is a fixed point of type $(0,1)$, and the dimension of the irrelevant directions indeed agrees with 
the formula in \eq{eq:Drelev}.
We can also explicitly see that the flow goes along the first-order phase transition lines, that is consistent with 
Section~\ref{sec:firstorder}.

There is a singular locus of the flow near $x_1\sim 0.6$. This is a singular locus of taking the gauge 
\eq{eq:gfix} \cite{Sasakura:2014zwa,Sasakura:2015xxa}, and has no physical relevance.

\subsection{A cyclic RG-flow in $N=3$} 
\label{sec:cyclicex}
In this subsection we will provide an explicit example of the cyclic flow discussed in Section~\ref{sec:deg}. Let us consider 
a $(0,2)$ fixed point for $N=3$:
\[
P=\frac{8 P\bphi^3}{27} \bphi\otimes \bphi \otimes \bphi - \frac{4 P\bphi^3}{3}[ \bphi\otimes I^-].
\]
As was discussed above \eq{eq:btphi},  $P$ is located on a first-order phase transition surface 
between the phases with $\bphi$ and $\tphi= -\bphi/2+ \frac{\sqrt{3}|\bphi|}{2} \eta_-\ (\eta_-\in V_-,\ |\eta_-|=1)$.
Now let us add a small perturbation to $P$, which corresponds to an infinitesimal jump\footnote{Though the small addition
is finite for an explicit example, the size can be made infinitely small, justifying ``infinitesimal".}:
\[
\tilde P=P+\epsilon\, \tphi \otimes \tphi \otimes \tphi,
\]
where $\epsilon$ is a small positive number. Then $\tphi$ becomes the unique ground state of $\tilde P$.
By checking the eigenvalues of $\tilde R^{\perp}=\tilde P \tphi/\tilde P \tphi^3$, one can find that 
this addition makes a jump,
\[
(0,2) \Rightarrow (0,0) \hbox{ close to } (1,1),
\]
which corresponds to $(N_+,N_-)=(0,2), (\tilde N_+, \tilde N_-)=(0,0)$ in \eq{eq:particularcyclic}.
Then it goes back as
\[
 (0,0) \rightarrow (0,2)
\]
by a continuous flow in the phase with $\tphi$.

Let us make a comment. By numerically solving the flow equation \eq{eq:Pflow} in the above process, 
one can find that the values of the initial $P$ and the final $P_{\rm final} $ are different, 
though they have the same type $(0,2)$. This is because they have different values of the ground states, $\bphi$ and $\tphi$,
respectively. This is just a difference of a gauge transformation. In fact $\bphi$ and $\tphi$ can be transformed by a $2 \pi/3$ rotation, and
by performing the same rotation on $\tilde P_{\rm final}$, one can find that they coincide (almost identically, 
because the process contains a small jump). Therefore more precisely the cyclic flow should be written as
\[
P\Rightarrow \tilde P \rightarrow P_{\rm final} \xrightarrow[]{\rm Gauge\, Trans.} P.
\]

\begin{figure}
\begin{center}
\includegraphics[width=7cm]{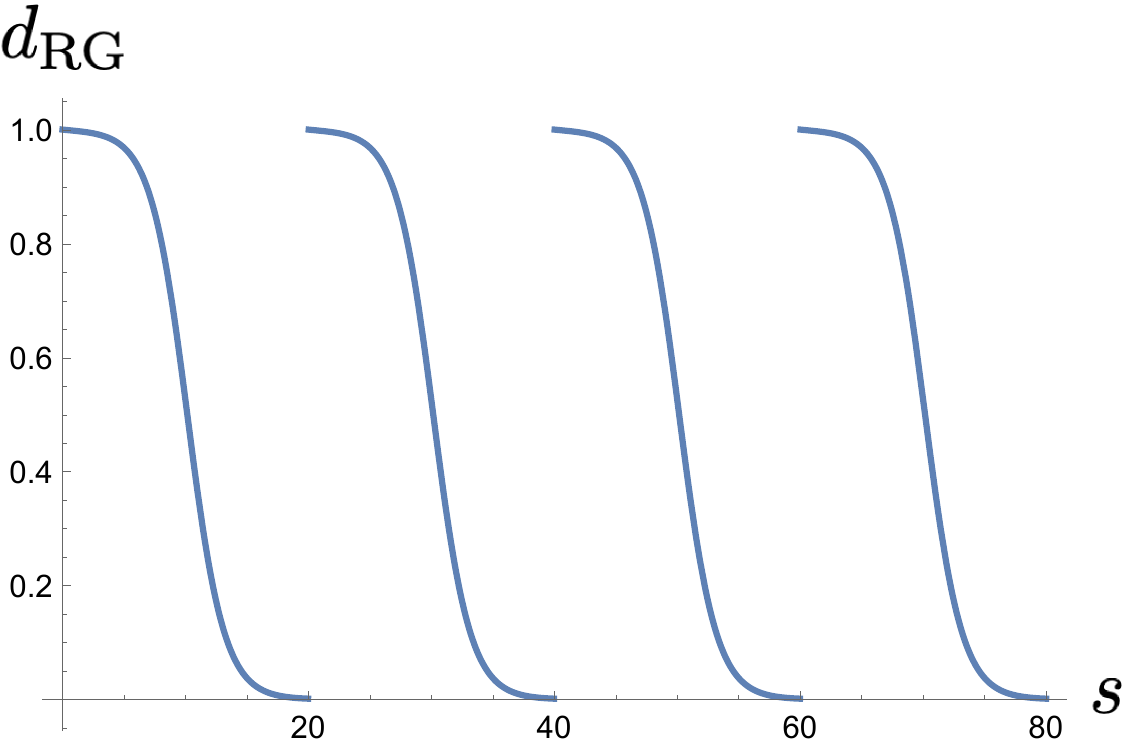}
\caption{The development of the RG-function for the cyclic flow. The figure is shown for its four cycles. }
\label{fig:drg}
\end{center}
\end{figure}

On the other hand, one can study the development of the RG-function $d_{\rm RG}(s)$ without the complication of gauge redundancy, 
because the RG-function is gauge invariant.  
In Figure~\ref{fig:drg} the development of the RG-function $d_{\rm RG}(s)$ for this example is shown for four cycles.  
It makes an increasing jump with the infinitesimal jump of $P$.

\section{Summary and future prospects}
In this paper we have established the connection between the dynamics of the randomly connected tensor networks (RCTN) with a real tensor 
and the canonical tensor model (CTM), which is a model of quantum gravity. 
A Hamiltonian vector flow using the Hamiltonian of CTM including the positive cosmological constant term 
has been shown to be what would be a renormalization group (RG) flow in RCTN. 
We have shown various properties of the flow, which support this identification.
Considering that RCTN does not have a fixed lattice structure and hence no renormalization procedures are expected to exist, 
the presence of such a flow seems stimulating. 
We have shown that the flow equation can be derived by taking the thermodynamic limit of an identity satisfied by the partition function
of RCTN.
We have studied the properties of the fixed points and have provided their complete classification.
We have shown that there exists an RG-function which monotonically decreases under the flow, resembling the $a$- and $c$- functions 
\cite{Zamolodchikov:1986gt,Cardy:1988cwa} in quantum field theories. 
On the other hand we have pointed out the presence of cyclic flows, if infinitesimal jumps near the fixed points are allowed. 

Similar connections between spacetime developments in quantum gravity and renormalization group flows exist in holographic 
settings \cite{deBoer:1999tgo,Strominger:2001pn}. While the RCTN with a real tensor, which we have studied in this paper,  
is too simple to become a sensible model of emergent spacetimes, 
the concrete explicit results of this paper would be useful as a basis to further pursue such connections in quantum gravity.
Since renormalization processes make assumed fundamental discreteness invisible, such connections would naturally lead to 
the idea of emergent continuous spacetimes in quantum gravity. 
It would be interesting to study this for RCTN with more general types of tensors (for instance complex ones), 
and also for the wave function of 
CTM, whose spacetime interpretation was partially studied in \cite{Kawano:2018pip,Kawano:2021vqc}.

Though the properties of the flow show some supportive evidences as an RG flow, it is not clear how we can link the flow to
a coarse graining process as in quantum field theories\footnote{But see a suggestive argument in \cite{Sasakura:2015xxa}.}. In particular the presence of the cyclic flows by allowing infinitesimal 
jumps from one phase to another may cast some doubts on a possible interpretation as a coarse graining process, since there 
are sudden transitions between large and small scales.
On the other hand we have a good example of an exchange of small and large scales, namely, the T-duality \cite{tduality}
in string theory. Therefore a stimulating interpretation of the presence of the cyclic flows is 
that small and large scales can be exchanged in quantum gravity, 
and RCTN may be giving a simple example.  

The results of this paper would also be useful in different areas.
As explained in Section~\ref{sec:RCTN} the stationary point equation is the same as
the tensor eigenvalue/vector equation \cite{Qi,lim,cart,qibook}. Solving it is known to be NP-hard \cite{nphard}, 
and the properties of the tensor eigenvalues/vectors are far from being well understood. In particular the largest tensor eigenvalue
is related to various applications,
such as the geometric measure of entanglement \cite{shi,barnum} in quantum information theory. A unique view point of this paper 
is that the tensor eigenvalue/vector problem is combined with the flow. In fact, some of the results in this paper 
are only based on the stationary point solution but not on the solution being a ground state.
This means that some of the results of this paper can also be applied to the tensor eigenvalue/vector problem, and would be useful 
to get some hints to its global properties. 

\vspace{.3cm}
\section*{Acknowledgements}
%%%%%%%%%%%%%%%%%%%%%%%%%%%%%%%%%%%%%%%%%%%%%%%
%\centerline{\bf Acknowledgements} 
The author would like to thank N.~Delporte for some correspondences.
The author was supported in part by JSPS KAKENHI Grant No.19K03825 and No.25K07153.  
%%%%%%%%%%%%%%%%%%%%%%%%%%%%%%%%%%%%%%%%%%%%%%%%

\appendix
\def\thesection{Appendix  \Alph{section}}
%\counterwithin{equation}{section}
%\renewcommand{\theequation}{\Alph{section}\arabic{equation}}
\section{Proof of \eq{eq:genP1}}
\label{app:genP1}
First of all, by the decomposition $V=V_\parallel\oplus V_\perp$, $P$ can generally be decomposed into the form, 
\[
P=a \, \bphi\otimes \bphi\otimes \bphi+\left[ \bphi\otimes \bphi \otimes v^\perp\right] +\left[ \bphi \otimes w^\perp\right] +P^\perp,
\label{eq:pstart}
\] 
where $a$ is real, $v^\perp\in V_\perp$ and $w^\perp \in [V_\perp \otimes V_\perp]$.
Contracting $P$ with $\bphi$ leads to
\[
P\bphi= \frac{3a}{2}  \bphi\otimes \bphi +\left[\bphi \otimes v^\perp \right] +\frac{1}{2} w^\perp, 
\]
where we have taken care of the symmetric factor as in \eq{eq:symmetrization} and the normalization \eq{eq:phi2}. By comparing this with 
\eq{eq:defofR} and \eq{eq:Rdecomp}, we obtain
\[
a=\frac{8}{27} P\phi^3,\ v^\perp=0,\ w^\perp=2 (P\phi^3) R^\perp.
\label{eq:avp}
\]
Plugging these into \eq{eq:pstart}, we obtain \eq{eq:genP1}.

\section{Proof of \eq{eq:pvanish}}
\label{app:pvanish}
The proofs are similar to that for \eq{eq:eineq}. We will find a $\phi\ (|\phi|^2=3/2)$ which satisfies $P\phi^3>P\bphi^3$, unless \eq{eq:pvanish} 
is satisfied. Here the presence of such $\phi$ would contradict the definition of $\bphi$, which maximizes $P\phi^3$ (See \eq{eq:newfree}).

\subsection{$P^{+++}=0$}
Suppose $P^{+++}$ is a non-zero tensor. Then there exists\footnote{One can check that 
$P(\eta_1+\eta_2+\eta_3)^3- P(\eta_1+\eta_2-\eta_3)^3-P(\eta_1-\eta_2+\eta_3)^3-P(-\eta_1+\eta_2+\eta_3)^3=24 P\eta_1\eta_2 \eta_3$ 
for any $\eta_i$. Therefore, if $P \eta^3$ identically vanishes, $P\eta_1 \eta_2 \eta_3$ identically vanishes. Therefore if $P$ is a
non-zero tensor, then there exists at least one $\eta$ which satisfies $P\eta^3\neq 0$. By taking an appropriate sign of $\eta$, we find
$\eta$ with $P\eta^3>0$.} 
at least one vector $\eta_+ \in V_+$ with $P^{+++} \eta_+^3>0,\ |\eta_+|=1$. 
Then consider $\tphi_\theta=\bphi \cos \theta+\eta_+ |\bphi| \sin\theta$, which has the norm 
$|\tphi_\theta|^2=3/2$. We obtain
\[
P\tphi_\theta^3&=P\bphi^3 \left(
\cos^3 \theta +\frac{3}{2} \cos \theta \sin^2 \theta +\frac{|\bphi|^3}{P\bphi^3} P^{+++}\eta_+^3 \sin^3\theta
\right) \cr
&=P\bphi^3 \left(1+\frac{|\bphi|^3}{P\bphi^3} P^{+++}\eta_+^3 \theta^3+O(\theta^4)\right),
\label{eq:pppp}
\] 
where we have put \eq{eq:pdecomp} and expanded in $\theta$ around $\theta=0$. 
Note that the other components of $P$ such as $P^{++-}$, etc., do not contribute,
because $P$ is contracted only with $\bphi$ and $\eta_+$. 
\eq{eq:pppp} shows that there exists $\theta\sim 0$ which satisfies $P\tphi_\theta^3>P\bphi^3$. This is a contradiction.

\subsection{$P^{---}=0$}
Suppose $P^{---}$ is a non-zero tensor. Then there exists at least one $\eta_-\in V_-$ with $P^{---} \eta_-^3>0,\ |\eta_-|=1$. 
Then consider $\tphi_\theta=\bphi \cos \theta+\eta_- |\bphi| \sin\theta$, which has the norm 
$|\tphi_\theta|^2=3/2$. We obtain
\s[
P\tphi_\theta^3&=
P\bphi^3 \left(
\cos^3 \theta -3 \cos \theta \sin^2 \theta +\frac{|\bphi|^3}{P\bphi^3} P^{---}\eta_-^3 \sin^3\theta
\right) \\
&=P\bphi^3 \left(1+\frac{|\bphi|^3}{P\bphi^3} P^{---}\eta_-^3 \left(\frac{\sqrt{3}}{2}\right)^3
\right) \\
&>P\bphi^3, 
\s]
where we have put $\theta=2 \pi/3$. This is a contradiction.

\subsection{$P^{--+}=P^{--\pp}=0$}
Suppose $P^{--+}=P^{--\pp}=0$ is not satisfied. Then one can find at least one pair, $\eta_- \in V_-$ and $\eta_ {+\pp} \in V_+\oplus V_\pp$ 
which satisfy $(P^{--+}+P^{--\pp})\eta_-^2 \eta_{+\pp}>0$. We can further assume $|\eta_-|=|\eta_{+\pp}|=1$. 
Then let us consider $\tphi_{\theta,\varphi}=\bphi \cos \theta +\eta_{-} |\bphi| \sin\theta \cos \varphi+\eta_{+\pp}|\bphi| \sin\theta \sin\varphi$, 
which has the norm $|\tphi_{\theta,\varphi}|^2=3/2$. 
Then, for $\theta=2 \pi /3$ and $\varphi\sim 0$, we obtain 
\[
P\tphi_{\theta,\varphi}^3=P\bphi^3 \left( 1+3 \left(\frac{\sqrt{3}}{2}\right)^3 \frac{(P^{--+}+P^{--\pp})\eta_-^2 
\eta_{+\pp} |\bphi|^3}{P\bphi^3} \varphi 
+O(\varphi^2)\right),
\]
where we have used the former result $P^{---}=0$.
Therefore there exists $\varphi\sim 0$ which satisfies $P\tphi_{\theta,\varphi}^3>P\bphi^3$. This is a contradiction.

\section{Computation of the second term in \eq{eq:genflowphi}}
\label{app:compsecond}
By explicit computation we obtain
\s[
\frac{\partial^2 f(P,\bar \phi)}{\partial \bar \phi_d\partial P_{abc}}&=-\frac{1}{P\bphi^3} 
\left( \delta_{ad} \bphi_b \bphi_c+\delta_{bd} \bphi_c \bphi_a+\delta_{cd} \bphi_a \bphi_b\right)
+\frac{3 (P\bphi^2)_d}{(P\bphi^3)^2} \phi_a \phi_b \phi_c \\
&= -\frac{1}{P\bphi^3}\left( 
\delta_{ad} \bphi_b \bphi_c+\delta_{bd} \bphi_c \bphi_a+\delta_{cd} \bphi_a \bphi_b
-2 \bphi_a \bphi_b \bphi_c \bphi_d
\right),
\label{eq:deroff}
\s]
where we have used the normalization of the derivative, 
$\frac{\partial}{\partial P_{abc}} P_{def} =\frac{1}{6} \sum_\sigma \delta_{a \sigma_d} \delta_{b \sigma_e} 
\delta_{c \sigma_f}$\footnote{This normalization of the derivative is necessary to 
realize $\delta P_{abc} \frac{\partial}{\partial P_{abc}}$ with a correct weight, which is used in the discussions.
For instance, we correctly obtain $\delta P_{abc} \frac{\partial}{\partial P_{abc}} (P_{def}P_{def})=2 P_{abc}\delta P_{abc}$.
This definition is also employed in \eq{eq:defofder}.},
and have applied \eq{eq:simple} to the last term in the first line. 
Let us denote $A_{abcd}\equiv\delta_{ad} \bphi_b \bphi_c+\delta_{bd} \bphi_c \bphi_a+\delta_{cd} \bphi_a \bphi_b$. Then let us
compute its contraction with the first term $[\bphi PP]$ in \eq{eq:Pflow}:
\s[
A_{abcd} [\bphi PP]_{abc}&=3 [\bphi PP]_{dbc} \bphi_b \bphi_c \\
&= 3 (P\bphi)_{da} (P\bphi^2)_a \\
&=\frac{4}{3} (P\bphi^3)^2 \bphi_d,
\s]
where we have used \eq{eq:simple} twice. On the other hand, as for the second term in \eq{eq:deroff}, we obtain
\s[
\bphi_a  \bphi_b \bphi_c [\bphi PP]_{abc}=(P\bphi^2)^2=\frac{2}{3} (P\bphi^3)^2,
\s]
where we have used \eq{eq:simple}. Therefore the contraction of \eq{eq:deroff} with $[\bphi PP]$ vanishes. 

Let us next consider the second term of \eq{eq:Pflow}. We obtain
\[
A_{abcd} [\bphi\otimes I ]_{abc}=3 \bphi^2 \bphi_d=\frac{9}{2} \bphi_d,
\]
where we have used \eq{eq:phi2}. As for the second term of \eq{eq:deroff}, 
\[
\bphi_a \bphi_b \bphi_c [\bphi\otimes I]_{abc}=(\bphi^2)^2 =\frac{9}{4}.
\]
Therefore the contraction of \eq{eq:deroff} with $[\bphi\otimes I]$ vanishes. 
Thus it has been proven that the second term of \eq{eq:genflowphi} identically vanishes. 

\section{Sufficiency of \eq{eq:maxppm}}
\label{app:maxppm}
We want to prove that \eq{eq:maxppm} is the sufficient condition for the content of the parentheses not to exceed 1 in \eq{eq:ppt3}.  
Rather than using the trigonometric functions, we replace $\cos \theta$,$\sin \theta \cos \varphi$, $\sin\theta \sin\varphi$ 
with real variables $x,y,z$, respectively, and impose $x^2+y^2+z^2=1$. 
By introducing the Lagrange multiplier $l$ for the constraint, the content of the parentheses can be rewritten as
\[
h=x^3 + \frac{3}{2} x y^2 - 3 x z^2 + 3 a y^2 z - l (x^2 + y^2 + z^2-1),
\label{eq:defofh}
\]
where $a=P^{++-} \eta_+^2 \eta_ - |\bphi|^3/P\bphi^3$. For $a^2\neq 3/4$, the solutions to the stationary 
condition of $h$ with respect to $x,y,z$ can readily be obtained as 
\[
(x,y,z)=l \left(\frac{2}{3},0,0 \right),l \left( -\frac{1}{3},0,\pm \frac{1}{\sqrt{3}}\right), l \left(0,\pm \frac{\sqrt{2}}{3a} ,\frac{1}{3a}\right).
\label{eq:crixyz}
\]
After the normalization $x^2 + y^2 + z^2=1$, we obtain 
\[
(x,y,z)=p (1,0,0), p\left(-\frac{1}{2},0,\pm \frac{\sqrt{3}}{2}\right), p\left(0,\pm \frac{\sqrt{2}}{\sqrt{3}},\frac{1}{\sqrt{3}}\right), 
\label{eq:normxyz}
\]
where $p=\pm1$. Then by putting them to \eq{eq:defofh}, we obtain
\[
h=p ,p, \frac{2 a  p}{\sqrt{3}},
\]
respectively.  
In the case of $a=\pm \sqrt{3}/2$, the discretely located solutions \eq{eq:normxyz} get connected, and form a continuous solution,
\[
(x,y,z)=p \left( 1-2 a z, \pm 2 \sqrt{(a-z) z}, z \right)
\]
with $h=p$. Therefore, for $h\leq 1$ to hold, $|a|\leq \sqrt{3}/2$. 
This derives \eq{eq:maxppm}\footnote{Taking the absolute value on the lefthand side is not essential in \eq{eq:maxppm}, 
because its sign can be flipped by $\eta_-\rightarrow -\eta_-$. }.

\section{Proof of the second statement}
\label{app:first}
We take the similar starting point as in \ref{app:maxppm}, but with a small real perturbation parameter denoted by $b$:
\[
h_b=x^3 + \frac{3(1+b)}{2} x y^2 - 3 x z^2 + 3 a y^2 z
\label{eq:defofhwithb}
\]
with  $x^2+y^2+z^2=1$. We assume $|a|<\sqrt{3}/2$, corresponding to \eq{eq:strict}. 

Let us first consider the effect of the perturbation around the solution $(x,y,z)=(1,0,0)$, which is the first solution with $p=1$ in \eq{eq:normxyz}.
By putting $x=\sqrt{1-y^2-z^2}$ to \eq{eq:defofhwithb} and checking the order of $y,z$ in $b$, we can find $y\sim O(\sqrt{b})$
and $z\sim O(b)$. More explicitly, by expanding \eq{eq:defofhwithb} in $b,y,z$ taking into account the orders in $b$ of $y,z$, we 
obtain
\[
h_b \sim  1+\frac{3 b y^2}{2} +3 a y^2 z -\frac{9 z^2}{2}-\frac{3 y^4}{8}+O(b^3).
\label{eq:hbexpan1}
\]
For $b<0$, the maximum of $h$ is at $(y,z)=(0,0)$, which corresponds to the original solution. 
However, for $b>0$, it splits into two maxima at
\[
(y,z)=\left( \pm \frac{\sqrt{6b}}{\sqrt{3-4 a^2}}, \frac{2 a b}{3-4 a^2}\right)
\]
with $h_b=1+9 b^2/(6-8 a^2)+O(b^3)$.

Let us next see the effect to the other two solutions in \eq{eq:normxyz}.
Let us expand $h_b$ around the solutions $(-\frac{1}{2},0,\pm \frac{\sqrt{3}}{2})$.
In this case we put $x=-\sqrt{1-y^2-z^2}$ and introduce a small perturbation $\delta z$ as $z=\pm \frac{\sqrt{3}}{2}+\delta z$. 
Checking the orders of $y,\delta z$ shows that they are $O(1)$ in $b$. In fact, expanding $h_b$, we obtain  
\[
h_b\sim 1-\frac{9}{4}\left( 1 \mp \frac{2 a }{\sqrt{3}} \right) y^2 -18\,( \delta z)^2 +O(b),
\]
which has the maximum at $(y,\delta z)=(0,0)$ for $|a|<\sqrt{3}/{2}$. Namely, the perturbation of $b$ does not affect these solutions
essentially.

The above analysis shows that the critical fixed point exists on a boundary of the first-order phase transition surfaces, where
two of the phases merge. 

\section{Fixed points of $\tilde H$ being those of $\bar H$}
\label{app:fixedpt}
Suppose that there is a fixed point $P$ of $\tilde H$:
\[
\{\bar H + r {\cal D} +w_{ab} {\cal J}_{ab} ,P\}=0.
\label{eq:fixedth}
\]
Let us employ the general expression \eq{eq:genP1}. From \eq{eq:Pflow}, \eq{eq:leftdp}, \eq{eq:renr}, and \eq{eq:renp},
we find
\[
\{\bar H, P\}=\frac{4 (P\bphi^3)^2}{3} \left[\bphi\otimes  \left( R^\perp-\frac{1}{3} I^\perp  \right) \left( R^\perp+\frac{2}{3} I^\perp  \right) \right]
+P\bphi^3 \left[R^\perp P^\perp \right].
\label{eq:poissondhp}
\]
We also have 
\s[
\{{\cal D},P\}&=P=\frac{8 (P\bphi^3)}{27}\bphi\otimes \bphi \otimes \bphi + 2 (P\bphi^3) \left[\bphi\otimes R^\perp\right]+P^\perp,
\label{eq:poissondp}
\s]
and 
\s[
\{ w_{ab} {\cal J}_{ab}, P \}&=-[wP] \\
&=-\frac{8 P\bphi^3}{27} [ (w \bphi) \otimes \bphi \otimes \bphi ] 
- \frac{2 P\bphi^3}{3} [ (w \bphi) \otimes R^\perp] -\frac{4 P\bphi^3}{3} [\bphi \otimes (w R^\perp) ] -[ w P^\perp],
\label{eq:poissonjp}
\s]
where $(w \bphi)_a=w_{ab} \bphi_b$, $(w R^\perp)_{ab}=w_{ac}R^\perp_{cb}$, and $(wP)_{abc}=w_{ad} P_{dbc}$.

Putting  \eq{eq:poissondhp}, \eq{eq:poissondp}, and \eq{eq:poissonjp} into \eq{eq:fixedth}, 
and noting that $w \bphi$ and $\bphi$ are transverse due to $w_{ab}=-w_{ba}$, 
we find that the $\bphi\otimes \bphi\otimes \bphi$ term in \eq{eq:poissondp} cannot be canceled in \eq{eq:fixedth}. 
Therefore we must have $r=0$ for \eq{eq:fixedth} to be satisfied.  

For further discussions, it is convenient to express $M_{abc}$ under the Poisson algebra as a derivative operator $D_{abc}$
introduced in \eq{eq:defofder}. Then the condition \eq{eq:fixedth} with $r=0$ can be rewritten as 
\[
\left( \hat{ \bar H}+w_{ab} \hat {\cal J} _{ab} \right)   P=0,
\label{eq:oppcond}
\]
where $\hat{ \bar H}$ and $\hat {\cal J} _{ab}$ are the operators which are obtained by replacing $M_{abc}$ with $D_{abc}$
in $\bar H$ and ${\cal J} _{ab}$, respectively. 
From \eq{eq:oppcond} and  $[\hat{ \bar H}, w_{ab} \hat {\cal J} _{ab}]=0$, we obtain
\[
e^{s \hat{ \bar H}}  P =e^{-s\, w_{ab} \hat {\cal J} _{ab}} P
\]
for arbitrary $s$.
The lefthand side is nothing but the solution to the flow equation, and it asymptotically approaches to a fixed point 
in the $s\rightarrow \infty$ limit, 
as was proven in Section~\ref{sec:flow}. 
On the other hand, the righthand side remains oscillatory in $s\rightarrow \infty$, unless
\[
w_{ab} \hat {\cal J} _{ab} P=0,
\]
because $e^{-s\, w_{ab} \hat {\cal J} _{ab}}$ is an $SO(N)$ transformation.
This in turn concludes 
\[
 \hat{\bar H}  P =0
\] 
from \eq{eq:oppcond}. Therefore $P$ is a fixed point of the flow equation generated by $\bar H$.

\end{document}